\documentclass[a4paper,fleqn,usenatbib]{mnras}
\usepackage{bm}

\usepackage{graphicx}	

\usepackage[utf8]{inputenc}

\usepackage{float}
\usepackage{amsmath,amssymb}
\usepackage{color}
\usepackage{dcolumn}
\usepackage{geometry}

\usepackage{url}
\usepackage{epstopdf}

\newcommand{\Dh}[1]{${\mathcal D}_{#1{\rm h}}$}

\newcommand{\ortho}{\textit{ortho}}
\newcommand{\para}{\textit{para}}

\newcommand{\hcch}{$^{12}$C$_2$H$_2$}

\newcommand{\nhcch}{C$_2$H$_2$}
\newcommand{\cm}{cm$^{-1}$}

\newcommand{\Marvel}{{\sc Marvel}}
\newcommand{\Trove}{{\sc TROVE}}
\newcommand{\trove}{{\sc TROVE}}
\newcommand{\ai}{{\it ab initio}}

\title[ExoMol line lists - XXXVII: C$_2$H$_2$]{ExoMol molecular line lists - XXXVII: spectra of acetylene}

\author[Katy L. Chubb et al.]{
	Katy L. Chubb,$^{1,2}$\thanks{E-mail: katy.chubb.14@ucl.ac.uk, k.l.chubb@sron.nl}
	Jonathan Tennyson,$^{1}$\thanks{E-mail: j.tennyson@ucl.ac.uk}
	Sergey N. Yurchenko$^{1}$\thanks{E-mail: s.yurchenko@ucl.ac.uk}
	\\
	$^{1}$Department of Physics and Astronomy, University College London, London, WC1E 6BT, UK\\
	$^{2}$SRON Netherlands Institute for Space Research, Sorbonnelaan 2,
	3584 CA, Utrecht, Netherlands\\
}

\begin{document}
	\label{firstpage}
	\pagerange{\pageref{firstpage}--\pageref{lastpage}}
	\maketitle
	
	\begin{abstract}
		A new ro-vibrational line list for the ground electronic state
of the main isotopologue of acetylene, \hcch, is computed as part of the
ExoMol project. The aCeTY line list covers the transition wavenumbers
up to 10~000~\cm\  ($ \lambda >1$~$\mu$m),
with lower and upper energy levels up to 12~000~\cm\ and 22~000~\cm\
considered, respectively. The calculations are performed up to a maximum value
for the vibrational angular momentum, $K_{\rm max}=L_{\rm max}$~=~16, and
maximum rotational angular momentum, $J$~=~99. Higher values of $J$ were
not within the specified wavenumber window. The aCeTY line list is considered to
be complete up to 2200~K, making it suitable for use in characterising
high-temperature exoplanet or cool stellar atmospheres. Einstein-A coefficients,
which can directly be used to calculate intensities at a particular temperature,
are computed for 4.3~billion (4~347~381~911) transitions
between 5~million (5~160~803) energy levels. We make comparisons against
other available data for \hcch, and demonstrate this to be the most complete
line list available. The line list is available in electronic form from the
online CDS and ExoMol databases.
	\end{abstract}
	
	
	
	
	\section{Introduction}

	In its electronic ground state, acetylene, HCCH, is a linear tetratomic unsaturated hydrocarbon whose
spectra is important in a large range of environments. On Earth, these range
from the hot, monitoring of oxy-acetylene flames which are widely used for
welding and related activities \citep{12Gaydon.C2H2,10ScVaMe.C2H2}, to the
temperate, monitoring of acetylene in breath, giving insights into the nature of
exhaled smoke \citep{10MeScSk.C2H2}, vehicle exhausts \citep{10ScVaMe.C2H2}, and
other air-born pollutants~\citep{59HuGo.C2H2}. Acetylene is also important in
the production of synthetic diamonds using carbon-rich
plasma~\citep{12KeRiWe.C2H2}.

	Further out in our solar system, acetylene is found in the atmospheres
of cold gas giants Saturn \citep{00MoBeLe.exo,97DeFeBe.planets}, Uranus
\citep{86EnCoAt} and Jupiter~\citep{74Ridgway.C2H2,86DrBeAt.C2H2}, the
hydrothermal plumes of Enceladus \citep{06WaCoIp,14MiBaOr.C2H2}, and in the
remarkably early-earth-like atmosphere of Titan
\citep{17Hoxxxx,08OrVoxx.C2H2,16SiMcCo.C2H2,19DiPuFa.c2h2}, where there has even
been some speculation as to acetylene's role in potential non-earth-like
life~\citep{05McSmxx,11Loxxxx,08OrVoxx.C2H2,87BeDaxx.C2H2,04Bains.exo,
13SeBaHu.exoplanet} and reactions involving molecules of pre-biotic
interest~\citep{17Hoxxxx,11Loxxxx,08OrVoxx.C2H2}. It has been detected on comets
such as Hyakotake \citep{96BrToWe.C2H2}, Halley, and 67P/Churyumov-Gerasimenko
\citep{15LeAlBa}. Even further into the galactic neighbourhood, acetylene
appears in star forming regions
\citep{76RiHaKl.C2H2,98DiWrCe.c2h2,18RaCoGa.C2H2}, is speculated to be an
important constituent of clouds in the upper atmospheres of brown dwarfs and
exoplanets
\citep{jt693,jt578,13BiRiHe.dwarfs,16MaAgMo,13OpBaBe.exo,11ShFoGr.C2H4}, and is
thought to play an important role in dust formation \citep{14DhRaxx.C2H2} and
AGB star evolution and atmospheric composition
\citep{00JoHrLo.C2H2,04Cernicharo.C2H2,04GaHoJo,99LoHoJo,09ArGiNo}, providing a major source of opacity in cool carbon
stars~\citep{82RiBaRa.C2H2,04GaHoJo}. For example, \nhcch\ was detected in the
carbon star Y CVn by \cite{78GoBrSt.C2H2} and in the low-mass young stellar
object IRS~46 by \cite{05LaDiBo.C2H2}. The first analysis of the atmosphere of a
``super-Earth'' exoplanet, 55~Cancri~e by \cite{jt629}, speculates that acetylene could be present in its atmosphere; however the spectral
data available at the time did not allow for an accurate verification of its presence in such a high temperature environment. A similar conclusion was found for the ``hot Jupiter'' extrasolar planet HD 189733b \citep{14KoBiBr} and for carbon-rich stars in the Large Magellanic Cloud \citep{06MaWoSl.C2H2,09LeAr,09MaAr}.


The infra-red spectrum of acetylene has been well studied in the lab, see \cite{16AmFaHe.C2H2,17LyCaxx.C2H2}
for example; a complete, up to 2017, compilation of laboratory studies can be found in \citet{jt705}.  More recent studies include those of
\cite{18LyCa.c2h2,18CaPeHa.c2h2,19LyVaMo.HCCH,18LyBeHu.C2H2}.


	At the temperatures of many exoplanets and cool stars (up to around 3000 -- 4000~K \citep{07TaLeNi.exo,17GaStCo.exo}), molecules are expected in abundance \citep{86Tsuji.exo}.
	An essential component in the analysis of such astrophysical atmospheres is therefore accurate and comprehensive spectroscopic data for all molecules of astrophysical importance, for a variety of pressures and temperatures. While a large amount of highly accurate data have been determined experimentally for a number of such molecules, they have largely been measured at room-temperature and are thus not well suited to the modelling of high-temperature environments; theoretical data are required for this purpose. The ExoMol project \citep{jt528,jt631} was set up for this reason, to produce a database of computed line lists appropriate for modelling exoplanet, brown dwarf or cool stellar atmospheres.
	As a result, high quality variational line lists which are appropriate up
to high temperatures have been computed for a host of molecules as part of the
ExoMol project, including CH$_4$ \citep{jt564,jt572,jt698}, HCN/HNC
\citep{jt570}, NH$_3$ \citep{jt771}, PH$_3$ \citep{jt592}, H$_2$O$_2$ \citep{jt638}, SO$_2$
\citep{jt635}, H$_2$S \citep{jt640}, SO$_3$ \citep{jt641}, VO \citep{jt644},
CO$_2$ \citep{jt667}, SiH$_4$ \citep{jt701}, H$_2$O \citep{jt665}, C$_2$H$_4$
\citep{jt729}, and, as presented in this work, C$_2$H$_2$ (see also
\citet{jt730}). Other molecular spectroscopic databases include
HITRAN~\citep{HITRAN}, HITEMP~\citep{HITEMP2010}, CDMS~\citep{cdms},
GEISA~\citep{GEISA}, TheoReTS~\citep{TheoReTS}, SPECTRA~\citep{spectra},
PNNL~\citep{PNNL}, MeCaSDA and ECaSDa~\citep{13BaWeSu}; however none
of these provide line lists for hot acetylene.  The ASD-1000 database of \citet{17LyPe.C2H2}
provides data on acetylene transitions which is designed to be valid for temperatures
up to 1000 K; we compare with this database below.

	
	Acetylene is a four-atomic (tetratomic) molecule which is linear in its
equilibrium configuration. The rotation-vibration spectrum of a polyatomic
molecule of this size, at the temperatures of exoplanets and cool stars,
typically spans the infra-red region of the electromagnetic spectrum. In this
region, only transitions between rotation-vibration (ro-vibrational) levels are
important; electronic transitions are of too high energy to be of interest. Such
ro-vibrational calculations essentially require a solution to the nuclear-motion
Schr{\"o}dinger equation, with some approximations required to enable feasible
computational treatment. The challenge with acetylene comes with its linear
geometry at equilibrium structure; linear molecules require special
consideration for calculations of ro-vibrational energies.  This was
demonstrated by \citet{68Waxxxx.linear} and very recently by \citet{jt730}; these two
approaches differ in their choice of internal coordinates used to represent
the vibrational Hamiltonian.

   This paper is structured as follows.  In Section~\ref{sec:calc} we outline the details of the calculations used to produce the aCeTY line list. This Section includes details on the basis set in Section~\ref{sec:basis}, the potential energy surface (PES) in Section~\ref{pes}, the refinement of this surface to empirical energy levels in Section~\ref{refinement}, empirical band centre replacement in Section~\ref{sec:bandcentre}, and details of the dipole moment surface (DMS) and its subsequent scaling in Sections~\ref{dms} and~\ref{sec:dipole:scaling}, respectively. The results of the line list calculations are given in Section~\ref{linelist}, with comparisons of the resulting spectra made against previous works in Section~\ref{compare}. In Section~\ref{sec:exo}, we demonstrate the differences in applying different line list data to exoplanet atmosphere modelling.  We give our summary in Section~\ref{summary}.
	


\section{Calculations}\label{sec:calc}


The $(3N-5)$ model for treating a four-atomic linear molecule such as HCCH has been fully implemented in the variational nuclear motion program \trove\ (Theoretical ROVibrational Energies) \citep{TROVE,15YaYuxx.method,17YuYaOv.methods}, as detailed in \citet{jt730}. Here, we outline only the main calculation steps towards computing the extensive aCeTY ro-vibrational line list for \nhcch\ in its ground electronic state.

\subsection{Basis set}\label{sec:basis}

The polyad number used to control the size of the primitive and contracted basis sets is given by:
\begin{equation}\label{eq:12}
	P = 2n_1 + n_2 + n_3 + n_4 + n_5 + n_6 + n_7\leq P_{\rm max}.
\end{equation}
Here, the vibrational quantum numbers follows the \Trove\ basis set selection, with $n_1$ corresponding to the excitation of the C--C stretching mode, $n_2$ and $n_3$ representing the C--H$_1$ and C--H$_2$ stretching modes and $n_4$, $n_5$, $n_6$ and $n_7$ representing the bending modes (see Table~\ref{t:QN}). This local mode notation deviates from the standard normal mode quantum numbers used for \hcch, most notedly for the bending modes: the \Trove\ bending quantum umbers  $n_4$,  $n_5$, $ n_6$, $n_7$ represent excitations along the  $\Delta x_1$, $\Delta y_1$, $\Delta x_2$, $\Delta y_2$, while the corresponding normal mode quantum numbers correspond to symmetric ($\upsilon_4$) and asymmetric ($\upsilon5$) modes as well as to the corresponding vibrational angular momenta ($\ell_4$ and $\ell_5$), see Table~\ref{t:QN} and also Section~\ref{linelist}.

For a linear molecule such as HCCH, another condition has been introduced in \Trove\ to control the basis set size; a maximum value for the total vibrational angular momentum, $L_{\rm max}$, which is equal to the $z$-projection of the rotational angular momentum, $K_{\rm max}$. This is linked to the total number of bending mode quanta (i.e. $n_{\rm bend} = n_4 + n_5 + n_6 + n_7$) in each vibrational band. Therefore we have a condition that:
\begin{equation}\label{e:lmax_kmax}
L_{\rm max} = K_{\rm max} \leq n_{\rm bend (max)},
\end{equation}
which is linked to the polyad number of Eq.~\eqref{eq:12}.
The aCeTY line list is relatively small in comparison to other polyatomic molecules of this size, largely due to the fact that the $K = L$ condition limits the number of allowed rotational sub-states in a vibrational band.  As vibrational states go up quickly in energy with increasing $n_{\rm bend}$, their energy will also rise quickly with increasing values of $L$. We therefore do not expect high values of $L$ to contribute until much higher energies.

	\begin{table}
		\caption{Quantum numbers used to classify the energy states of acetylene, \hcch.}
		\label{t:QN}
		\footnotesize
			\begin{tabular}{ll}
				\hline\hline
				Label & Description\\
				\hline
             \multicolumn{2}{c}{Conventional (normal mode) quantum numbers} \\	
 			$\upsilon_1$   & CH symmetric stretch ($\Sigma_g^{+}$)        \\
				$\upsilon_2$   & CC symmetric stretch ($\Sigma_g^{+}$)        \\
				$\upsilon_3$   & CH antisymmetric stretch ($\Sigma_u^{+}$)    \\
				$\upsilon_4$   & Symmetric (trans) bend ($\Pi_g$)             \\
				$\ell_4$        & Vibrational angular momentum associated with $\upsilon_4$    \\
				$\upsilon_5$   & Antisymmetric (cis) bend ($\Pi_u$)        \\
				$\ell_5$        & Vibrational angular momentum associated with $\upsilon_5$ \\
				$L=|l|$ & Total vibrational angular momentum, $|\ell_4+\ell_5|$    \\
				$K=|k|$ & Rotational quantum number; $z$-projection of ${\bf J}$ \\
				$J$ & QN associated with rotational angular momentum, ${\bf J}$. \\
				$e/f$ & Rotationless parity of the ro-vibrational state \\
				\ortho/\para\   & Nuclear spin state, see \citet{18ChJeYu.C2H2} \\
\hline
\multicolumn{2}{c}{\Trove\ local mode quantum numbers} \\
$n_1$  &  CC symmetric stretch \\
$n_2$  &  CH$_1$ stretch       \\
$n_3$  &  CH$_2$ stretch      \\
$n_4$  &  $x_1$  bend         \\
$n_5$  &  $y_1$  bend         \\
$n_6$  &  $x_2$  bend         \\
$n_7$  &  $y_2$  bend         \\
$L=|l|$ & Total vibrational angular momentum  $L=K$ \\
$\Gamma_{\rm str}$ & Symmetry of the vibrational component (\Dh{n}, $n=34$) \\
$K=|k|$ & Rotational quantum number; $z$-projection of ${\bf J}$ \\
$\Gamma_{\rm rot}$ & Symmetry of the rotational component (\Dh{n}, $n=34$) \\
$J$ & QN associated with rotational angular momentum, ${\bf J}$. \\
$\Gamma_{\rm tot}$ & Symmetry of the rotational component (\Dh{n}, $n=34$) \\
				\hline\hline
			\end{tabular}
	\end{table}

\Trove\ uses a multi-step contraction scheme. At step 1, the stretching primitive basis functions $\phi_{n_1}(\xi_1)$, $\phi_{n_2}(\xi_2)$ and $\phi_{n_3}(\xi_3)$ are generated using the Numerov-Cooley approach~\citep{TROVE,24Numerov.method,61Cooley.method} as eigenfunctions of the corresponding 1D reduced stretching Hamiltonian operators $\hat{H}_{i}^{\rm (1D)}$, obtained by freezing  all other degrees of freedom at their equilibrium values in the $J=0$ Hamiltonian.  For the bending basis functions, $\phi_{n_4}(\xi_4),\ldots,\phi_{n_7}(\xi_7)$, 1D harmonic oscillators are used. These seven 1D basis sets are then combined into three sub-groups
	\begin{equation}\label{e:class1_2}
	\phi^{\rm (1D)}_{n_1}(\xi_1) = \phi_{n_1}(\xi_1),
	\end{equation}
	\begin{equation}\label{e:class2_2}
	\phi^{\rm (2D)}_{n_2 n_3}(\xi_2, \xi_3) = \phi_{n_2}(\xi_2) \phi_{n_3}(\xi_3),
	\end{equation}
	\begin{equation}\label{e:class3_2}
	\phi^{\rm (4D)}_{n_4 n_5 n_6 n_7}(\xi_4, \xi_5, \xi_6, \xi_7) = \phi_{n_4}(\xi_4) \phi_{n_5}(\xi_5) \phi_{n_6}(\xi_6) \phi_{n_7}(\xi_7)
	\end{equation}
and used to solve  eigenvalue problems for the three corresponding reduced Hamiltonian operators:  stretching $\hat{H}^{\rm (1D)}$ and $\hat{H}^{\rm (2D)}$, and bending $\hat{H}^{\rm (4D)}$. The reduced Hamiltonians $\hat{H}^{(N{\rm D})}$  ($N=1,2,4$) are constructed by averaging the total vibrational Hamiltonian operator $\hat{H}^{(J=0)}$ over the other ground vibrational basis functions \citep{18ChJeYu.C2H2,jt730}. The eigenfunctions of the three reduced problems $\psi^{\rm (1D)}_{\lambda_1}$, $\psi^{\rm (2D)}_{\lambda_2}$ and $\psi^{\rm (4D)}_{\lambda_3}$ are  contracted and classified according with the \Dh{n}(M) symmetry using the symmetrisation procedure by \citet{17YuYaOv.methods}  to form a symmetry-adapted 7D vibrational basis set as a product  $\psi^{\rm (1D)}_{\lambda_1} \psi^{\rm (2D)}_{\lambda_2} \psi^{\rm (4D)}_{\lambda_3}$. At step 2, the $(J=0)$ eigenproblem is solved using this contracted basis. The eigenfunctions of the latter are then contracted again and used to form the symmetry-adapted ro-vibrational basis set, together with the spherical harmonics representing the rotational part.

For the current work, the polyad number in Eq.~\eqref{eq:12} was chosen as $P_{\rm max}$=18 for the primitive basis set and reduced to 16 after the 1st contraction. Energy cutoffs of 60~000~\cm, 50~000~\cm\ and 22~000~\cm\  were used for the primitive, contracted and $(J=0)$-contracted basis functions, respectively. The ro-vibrational basis set was formed using the energy cutoff of 22~000~\cm. The energies computed using these cutoff values are better converged than those of the \ai\ room-temperature line list of \citet{jt730}.
The vibrational and rotational states are classified with the \Dh{n} representations ($n=34$) and the projections of the vibrational and rotational angular momenta, $L$ and $K$, respectively, with the constraint $K=L$. The maximum value for the total vibrational angular momentum, $K_{\rm max}=L_{\rm max}$, used to build the multidimensional basis sets, see Eq.~(\ref{e:lmax_kmax}), is 16.  The ro-vibrational states can only span the four irreducible representations of \Dh{34};  $A_{1g}$, $A_{2g}$, $A_{1u}$ and $A_{2u}$. For the vibrational basis set used ($L_{\rm max} = 16$), the symmetry group \Dh{34} is equivalent to \Dh{\infty}.
The following selection rules apply to the electric dipole transitions of \hcch:
\begin{eqnarray}
J' + J''  >0 \quad {\rm and} \quad   J' \leftrightarrow  J'' \pm 1, \\
A_{1g} \leftrightarrow  A_{1u} \quad {\rm and} \quad A_{2g} \leftrightarrow  A_{2u}.
\end{eqnarray}
The corresponding nuclear statistical weights $g_{\rm ns}$ are 1 and 3 for the $A_{1g}, A_{1u}$  and $A_{2g}, A_{2u}$ pairs of states, respectively.
The kinetic energy and potential energy expansions are truncated at 2$^{\rm nd}$ and 8$^{\rm th}$ order, respectively (the kinetic energy terms of higher than 2$^{\rm nd}$ order appear to contribute very little to the calculated ro-vibrational energies, with expansion to higher orders becoming more computationally demanding). The equilibrium bond lengths are set to  1.20498127~\AA\ and 1.06295428~\AA\ for the C-C and C-H bonds, respectively. Nuclear masses were used. Calculations were performed up to a high value of $J=99$, which was determined by the maximum values of lower and upper energies used in the line list calculations; these have an effect on the temperature dependence of the line list, as discussed in Section~\ref{linelist}.

\subsection{Potential Energy Surface}\label{pes}

\trove\ represents all components of the Hamiltonian operator using a Taylor expansion about the equilibrium structure in terms of the linearised coordinates $\xi_\lambda$, $\lambda=1 \ldots 7$  (or some 1D functions of them). This leads to a sum-of-product form which allows the the matrix to be
computed as products of  1D-integrals. In general the
potential energy surface (PES), $V(\bxi)$, is represented in terms of user-chosen curvilinear coordinates;  \trove\ uses a quadruple-precision numerical finite difference method to re-expand $V(\bxi)$ in terms of the \trove-coordinates $\bxi = \{ \xi_\lambda\}$. As detailed in \cite{jt730}, the TROVE linearised coordinates for \nhcch\ are selected as
	\begin{align*}\label{e:lin}
	\xi_1 &= \Delta R^{\rm lin}, & \xi_2 &= \Delta r^{\rm lin}_1,  &	\xi_3 &= \Delta r^{\rm lin}_2,   \\
   	\xi_4 &=  \Delta x_1, &\xi_5 &=  \Delta y_1,   &	\xi_6 &=  \Delta x_2, &	\xi_7 &=  \Delta y_2.
	\end{align*}
where $R^{\rm lin}, r^{\rm lin}_1$ and $r^{\rm lin}_2$ are based on the curvilinear, bond-length coordinates $R\equiv r_{\rm CC}$, $r_1 \equiv r_{{\rm CH}_1}$ and $r_2 \equiv r_{{\rm CH}_2}$; $x_1, y_1,  x_2$ and $ y_2$ are Cartesian coordinates of the hydrogen atoms along the $x$ and $y$ axes. The displacements are taken from the equilibrium values of $R$, $r_1$ and $r_2$, respectively. The equilibrium values (at the linear configuration) of $x_i$ and $y_i$ ($i=1,2$) are zero.

Here we use the potential energy function of \nhcch\ reported recently by \citet{jt730}. It is represented in terms of the linelarised coordinates as follows:
	\begin{equation}\label{e:PEF}
	V({\bf \chi}) = \sum_{i,j,k,\ldots} f_{i,j,k,\ldots} \chi_1^i \chi_2^j \chi_3^k \ldots ,
	\end{equation}
	where $\chi_\lambda$ are given by:
	\begin{eqnarray}\label{e:pes_lin}
	\chi_1 &=& 1-\exp\left(-a \,\Delta R^{\rm lin}   \right),   \\
	\chi_2 &=& 1-\exp\left(-b \,\Delta r^{\rm lin}_1 \right), \nonumber \\
	\chi_3 &=& 1-\exp\left(-b \,\Delta r^{\rm lin}_2 \right),  \nonumber\\
	\chi_4 &=&  \Delta x_1, \nonumber\\
	\chi_5 &=&  \Delta y_1, \nonumber \\
	\chi_6 &=&  \Delta x_2, \nonumber\\
	\chi_7 &=&  \Delta y_2. \nonumber
	\end{eqnarray}
Here $a$ and $b$ are two Morse parameters.
	

The \ai\ PES of \citet{jt730} was computed using MOLPRO \citep{MOLPRO} at the VQZ-F12/CCSD(T)-F12c level of theory \citep{08PeAdWe.ai}
on a grid of 66~000 points spanning the 6D nuclear-geometry coordinate space up to 50~000~\cm.
A least squares fit was used to determine the coefficients $f_{i,j,k,\ldots}$ in Eq.~(\ref{e:PEF}) to the \ai\ energies,
using a grid of 46~986 \ai\ points covering up to 14~000~\cm, with a weighted root-mean-square (\textit{rms}) error of 3.98~\cm\ and an un-weighted \textit{rms} of 15.65~\cm, using 358 symmetrised parameters expanded up to 8$^{\rm th}$ order.

To improve the accuracy of the variational calculations, here we refine the \ai\ PES of \nhcch\ by fitting the expansion potential parameters $ f_{i,j,k,\ldots}$ to experimental data,  as outlined below.

\subsection{Refinement of the Potential Energy Surface to Experimental Energy Levels}\label{refinement}
		
The refinement procedure is carried out under the assumption that the \ai\ PES can be used to initially determine a set of energy levels and eigenfunctions. In this case, a correction is added to the \ai\ PES in terms of a set of internal coordinates $\xi$ \citep{jt503}:
\begin{equation}\label{e:ref_correction}
	\Delta V =  \sum_{ijk\ldots }\Delta f_{ijk\ldots } \chi_1^i \chi_2^j \chi_3^k\ldots ,
\end{equation}
where $\Delta f_{ijk\ldots }$ are the refined parameters, given as correction terms to the expansion coefficients of the original PES in Eq.~\eqref{e:PEF}, with the symmetry of the molecule taken into account in the same way as for the original \ai\ PES. The eigenfunctions of the ``unperturbed'', \ai\ Hamiltonian are used as basis functions when solving the new ro-vibrational eigenproblems with the correction $\Delta V$ to the PES included. This process is performed iteratively in \Trove, with the fitting procedure making use of empirical energy levels which should be added in gradually, according to the level of confidence placed in them. For details of the \Trove\ refinement procedure the reader is referred to \citet{jt503}.
		
Highly accurate experimentally determined
data provide an essential component in the calculation of a high-quality line
list, for both effective Hamiltonian and the majority of variational approaches.
Fortunately for \hcch, a wealth of experimental ro-vibrational spectral data has
been recorded over the decades (see, for example,
\citet{17LyCaxx.C2H2,16AmFaHe.C2H2,07Herman.C2H2,10DiHexx.C2H2,11Herman.C2H2}).
\citet{jt705} gathered, collated and analysed all such experimental data from the literature
 for  \hcch. They used the  \Marvel\ (measured active vibration-rotation energy level) procedure
 \citep{jt412} to provide a set of empirically-derived energy levels.
We use these \Marvel\ energies  for the PES refinement procedure, with level weighted according to
the level of confidence in their experimental assignment. The \ai\ energies
which were used to fit the initial PES (see above) are also included in the
refinement procedure in order to constrain the shape of the refined PES to the
\ai\ PES (see \citet{03YuCaJe.PH3}).

Good quantum numbers for acetylene states are the rotational angular momentum quantum number $J$ and overall symmetry $\Gamma$. These are therefore the primary criteria used to match energy levels from the energy levels in the supplementary data of \citet{jt705} to those computed using \Trove. An important parameter in the \Marvel\ energy level output of \citet{jt705} is NumTrans, which gives the number of transitions linking a particular state to other energy levels. The higher the number of linking transitions, the higher the confidence which should be given to that empirical energy level. States with NumTrans=1 were deemed unreliable and were therefore not included into the fit.  A few states with NumTrans=2 and large residuals were also omitted.  The fact that vibrational states which include some quanta of C-H stretch are more likely to be observed in experiment than those without was used to inform our judgement when matching theoretical \Trove\ energy levels with their experimentally determined counterpart.
Table~\ref{t:obs-calc} is a summary of the fit, for which only experimental energies with $J\le 5$ were used; the  \textit{rms} errors between the experimental (\Marvel) and calculated (refined) ro-vibrational energies are shown for a set of vibrational bands in the column rms-I. The full Table is included as part of the supplementary information to this work.
The vibrational quantum numbers used for labelling the states of \hcch\ in Table \ref{t:obs-calc} are detailed in Table~\ref{t:QN} together with the rotational quantum numbers used typically (see \citet{jt705}).

The refined potential energy function is available as supplementary data to this article and from \url{exomol.com}.

\begin{table*}
\caption{An extract of the Obs.-Calc. residuals for \hcch. The root-mean-square errors between the experimental and calculated (refined) ro-vibrational energies for different vibrational bands (classified by their symmetry $\Gamma$, quantum numbers $\upsilon_1, \upsilon_2, \upsilon_3, \upsilon_4, l_4, \upsilon_5, l_5, k$ and energy $E_i/hc$), before (rms-I) and after (rms-II) the band-centre shifts. The full table is given as part of the supplementary information to this paper.  $E_i/hc$ is the \Trove\ energy after the band centre shift. }
\label{t:obs-calc}
\footnotesize
\begin{tabular}{lrrrrrrrrrrrr}
\hline\hline
 $\Gamma$             &$  \upsilon_1$  & $\upsilon_2$ & $\upsilon_3$ & $\upsilon_4$ & $l_4$ & $\upsilon_5$ & $l_5$ & $k$  & $E_i/hc$ & rms-I & rms-II  \\
\hline
 $ \Sigma_g^+   $&   0 &  0 &  0 &  0 &  0 &  0 &  0 &  0 &      0.00 &   0.0009 &      0.0009  \\
 $ \Pi_g        $&   0 &  0 &  0 &  1 &  1 &  0 &  0 &  1 &    612.86 &   0.0021 &      0.0021  \\
 $ \Pi_u        $&   0 &  0 &  0 &  0 &  0 &  1 &  1 &  1 &    730.33 &   0.0016 &      0.0016  \\
 $ \Sigma_g^+   $&   0 &  0 &  0 &  2 &  0 &  0 &  0 &  0 &   1230.38 &   0.0160 &      0.0042  \\
 $ \Delta_g     $&   0 &  0 &  0 &  2 &  2 &  0 &  0 &  2 &   1233.49 &   0.0151 &      0.0137  \\
 $ \Sigma_u^+   $&   0 &  0 &  0 &  1 &  1 &  1 & -1 &  0 &   1328.07 &   0.2978 &      0.0012  \\
 $ \Sigma_u^-   $&   0 &  0 &  0 &  1 &  1 &  1 & -1 &  0 &   1340.55 &   0.0045 &      0.0017  \\
 $ \Delta_u     $&   0 &  0 &  0 &  1 &  1 &  1 &  1 &  2 &   1347.51 &   0.0048 &      0.0048  \\
 $ \Sigma_g^+   $&   0 &  0 &  0 &  0 &  0 &  2 &  0 &  0 &   1449.11 &   0.0043 &      0.0012  \\
 $ \Delta_g     $&   0 &  0 &  0 &  0 &  0 &  2 &  2 &  2 &   1463.00 &   0.0082 &      0.0081  \\
 $ \Pi_g        $&   0 &  0 &  0 &  3 &  1 &  0 &  0 &  1 &   1855.77 &   0.0548 &      0.0037  \\
 $ \Pi_u        $&   0 &  0 &  0 &  2 &  2 &  1 & -1 &  1 &   1941.18 &   0.3933 &      0.0019  \\
 $ \Pi_u        $&   0 &  0 &  0 &  2 &  0 &  1 &  1 &  1 &   1960.87 &   0.0024 &      0.0023  \\
 $ \Sigma_g^+   $&   0 &  1 &  0 &  0 &  0 &  0 &  0 &  0 &   1974.35 &   0.0242 &      0.0185  \\
 $ \Pi_g        $&   0 &  0 &  0 &  1 &  1 &  2 &  0 &  1 &   2049.06 &   0.0298 &      0.0021  \\
 $ \Pi_g        $&   0 &  0 &  0 &  1 & -1 &  2 &  2 &  1 &   2066.97 &   0.0394 &      0.0031  \\
 $ \Phi_g       $&   0 &  0 &  0 &  1 &  1 &  2 &  2 &  3 &   2084.68 &   0.0708 &      0.0698  \\
 $ \Pi_u        $&   0 &  0 &  0 &  0 &  0 &  3 &  1 &  1 &   2170.34 &   0.1247 &      0.0005  \\
 $ \Sigma_u^+   $&   0 &  0 &  0 &  3 &  1 &  1 & -1 &  0 &   2560.59 &   0.1499 &      0.0027  \\
 $ \Pi_g        $&   0 &  1 &  0 &  1 &  1 &  0 &  0 &  1 &   2574.76 &   0.2551 &      0.0193  \\
 $ \Sigma_u^-   $&   0 &  0 &  0 &  3 &  1 &  1 & -1 &  0 &   2583.84 &   0.6489 &      0.0043  \\
 $ \Sigma_g^+   $&   0 &  0 &  0 &  2 &  2 &  2 & -2 &  0 &   2648.01 &   0.0233 &      0.0058  \\
 $ \Sigma_g^-   $&   0 &  0 &  0 &  2 &  2 &  2 & -2 &  0 &   2661.16 &   0.1688 &      0.0088  \\
 $ \Delta_g     $&   0 &  0 &  0 &  2 &  2 &  2 &  0 &  2 &   2666.06 &   0.5931 &      0.0009  \\
 $ \Pi_u        $&   0 &  1 &  0 &  0 &  0 &  1 &  1 &  1 &   2703.10 &   0.0073 &      0.0072  \\
 $ \Sigma_u^+   $&   0 &  0 &  0 &  1 &  1 &  3 & -1 &  0 &   2757.80 &   0.0314 &      0.0011  \\
 $ \Sigma_g^+   $&   0 &  0 &  0 &  0 &  0 &  4 &  0 &  0 &   2880.22 &   0.1967 &      0.0025  \\
 $ \Delta_g     $&   0 &  0 &  0 &  0 &  0 &  4 &  2 &  2 &   2894.04 &   0.0195 &      0.0161  \\
 $ \Sigma_u^+   $&   0 &  1 &  0 &  1 &  1 &  1 & -1 &  0 &   3281.91 &   0.0040 &      0.0040  \\
 $ \Sigma_u^+   $&   0 &  0 &  1 &  0 &  0 &  0 &  0 &  0 &   3294.85 &   0.0112 &      0.0039  \\
 $ \Sigma_u^-   $&   0 &  1 &  0 &  1 &  1 &  1 & -1 &  0 &   3300.65 &   0.0802 &      0.0078  \\
 $ \Delta_u     $&   0 &  1 &  0 &  1 &  1 &  1 &  1 &  2 &   3307.73 &   0.2140 &      0.0023  \\
 $ \Sigma_g^+   $&   1 &  0 &  0 &  0 &  0 &  0 &  0 &  0 &   3372.84 &   0.1657 &      0.0028  \\
 $ \Pi_u        $&   0 &  1 &  0 &  2 &  0 &  1 &  1 &  1 &   3882.42 &   0.9276 &      0.0032  \\
 $ \Pi_u        $&   0 &  0 &  1 &  1 &  1 &  0 &  0 &  1 &   3898.34 &   0.0481 &      0.0017  \\
 $ \Sigma_g^+   $&   0 &  2 &  0 &  0 &  0 &  0 &  0 &  0 &   3933.97 &   0.0387 &      0.0297  \\
 $ \Pi_g        $&   1 &  0 &  0 &  1 &  1 &  0 &  0 &  1 &   3970.05 &   0.0097 &      0.0043  \\
 $ \Pi_g        $&   0 &  1 &  0 &  1 &  1 &  2 &  0 &  1 &   4002.46 &   0.0396 &      0.0076  \\
 $ \Pi_g        $&   0 &  0 &  1 &  0 &  0 &  1 &  1 &  1 &   4016.73 &   0.0979 &      0.0056  \\
 $ \Pi_u        $&   1 &  0 &  0 &  0 &  0 &  1 &  1 &  1 &   4092.34 &   0.0558 &      0.0011  \\
 $ \Pi_u        $&   0 &  1 &  0 &  0 &  0 &  3 &  1 &  1 &   4140.08 &   0.7091 &      0.0084  \\
\hline
\hline
\end{tabular}
\end{table*}

\subsection{Band Centre Replacement and ``MARVELisation'}\label{sec:bandcentre}
		
As mentioned previously, \Trove\ uses a double layer contraction scheme with vibrational basis functions obtained as the solution of the $J=0$
problem. This $J=0$-representation has a more compact vibrational basis set and also facilitates the matrix elements calculations.
Indeed, the vibrational part of the ro-vibrational Hamiltonian is diagonal on this basis with the matrix elements given by the
corresponding vibrational ($J=0$) band centres $E_{i}^{(\rm vib)}$. An indirect advantage of this representation is a
direct access to the $J=0$ energies used in the consecutive ro-vibrational calculations allowing us to empirically modify the band
centres \citep{jt500}. This is a necessary procedure if the line list is to be used in any high-temperature, high-resolution Doppler-shift studies (see, for example~\cite{17BrLiBe.exo,14KoBiBr}),
where the line positions in a line list need to be as accurate as possible. In this work, 128 calculated band-centres were shifted to minimise the difference with the ro-vibrational \Marvel\ term values \citep{jt705}. Again, we only used \Marvel\ energies with $J\le 5$ based on more than one experimental data (NumTrans~>~1). A few NumTrans~=~2 states represented by large residuals were suspected as outliers and also were left out.  This has improved the accuracy of the resulting line positions, as demonstrated by Table~\ref{t:obs-calc}, which shows root-mean-square errors between the experimental and calculated (refined) ro-vibrational energies for a small number of vibrational bands, before (rms-I) and after the band-centre shifts (rms-II). The full Table is included as part of the supplementary information to this work. For details of the procedure see \citet{jt500}, where it was referenced as EBSC.
To improve the accuracy of our line positions, we ``MARVELise'' the data:
the energy levels in the ExoMol states file are replaced by the \Marvel\ energies of
\citet{jt705}.  The MARVEL uncertainties are kept as part of the line list.
 In doing this we take advantage of the ExoMol data format; see
\citet{jt548} for details. It should be noted that the \Marvel\ analysis was
performed in 2017, and a periodic update will be undertaken at some point in the
future in order to include experimental data which has been published since
then.		
We also provide indicative estimates of the uncertainties of all \Trove\ energies. This is done using the following approximation. 1) For all states which were replaced with \Marvel\ energies, we take the associated \Marvel\ uncertainty. 2) Where we have applied band centre shifting, we take the root-mean-square error for a particular band, before band shifting (i.e. rms-I in Table~\ref{t:obs-calc}).  3) For all other bands which have not been ``MARVELised'' or had a band centre shift applied, we use an approximate method to determine the shift, based on how the rms-I for each shifted band correlates with the number of quanta associated with the C-C stretch ($n_1$ in the local mode, \Trove\ notation), the C-H stretches ($n_2 + n_3$) and the bending modes ($n_4 + n_5 + n_6 + n_7$):
\begin{equation}\label{e:unc_approx}
0.4 n_2 + 0.5(n_1+n_3)+0.4(n_4+n_5)+0.003(n_4+n_5)^2.
\end{equation}
These uncertainities are then rounded to the nearest integer, with those under 0.5\cm\ rounded up to 0.5\cm. We stress that the estimation of these uncertainties in this way is very approximate, and the main aim is to distinguish between those states considered reliable and those states which should not be considered reliable when comparing to high-resolution observations.

\subsection{Dipole Moment Surface}\label{dms}
	
Here we use the \ai\ dipole moment surface (DMS)  computed by \citet{jt730} with the finite field method
in MOLPRO at the CCSD(T)/aug-cc-PVQZ level of theory on
a grid of 66~000 points  covering energies up to 50~000~\cm.
The electric dipole moment components, $\mu_\alpha$ ($\alpha$=$x,y,z$), were represented using the same set of seven linearised coordinates as for the $3N-5$ potential above to the following function:
	\begin{eqnarray}\label{e:dms_func}
	\mu_x ({\bf \zeta}) &=& \sum_{i} F_{i,j,k,\ldots}^x \zeta_1^{i} \zeta_2^{j} \zeta_3^{k}\ldots, \\
	\mu_y ({\bf \zeta}) &=& \sum_{i} F_{i,j,k,\ldots}^y \zeta_1^{i} \zeta_2^{j} \zeta_3^{k}\ldots, \\
	\mu_z ({\bf \zeta}) &=& \sum_{i} F_{i,j,k,\ldots}^z \zeta_1^{i} \zeta_2^{j} \zeta_3^{k}\ldots .
	\end{eqnarray}
	where $\zeta_\lambda$ are given by:
    \begin{align*}
    \zeta_1 &= \Delta R^{\rm lin}, & \zeta_2 &= \Delta r_1^{\rm lin},  &	\zeta_3 &= \Delta r_2^{\rm lin},  \\
    \zeta_4 &=  \Delta x_1, & \zeta_5 &=  \Delta y_1,  & 	\zeta_6 &=  \Delta x_2, & \zeta_7 &=  \Delta y_2.
    \end{align*}

Use was made of discrete symmetries (see \citet{18ChJeYu.C2H2}), and the three components of the dipole were expanded up to 7$^{\rm th}$ order and symmetrised according to the operations of \Dh{12}. The value of $n$ here in \Dh{n} is determined by the order up to which the function (dipole moment or potential energy) is expanded. See, for example \cite{ChubbPhD}. The three Cartesian components of the dipole moment, $\mu_x$, $\mu_y$, $\mu_z$, transform differently to one another ($\mu_x$ and $\mu_y$ as $E_{1u}$ and $\mu_z$ as $A_{2u}$ for $D_{nh}(M)$~\citep{06BuJexx.method}): the $\mu_x$ and $\mu_y$ components share the corresponding expansion parameters, while that the parameters for the $\mu_z$ component are independent.
This dipole moment function is provided as supplementary material to this work as a Fortran program.

\subsection{Vibrational Transition Dipole Moment Scaling }\label{sec:dipole:scaling}
	
In order to improve the quality of the line intensities, at least for the vibrational bands known experimentally from HITRAN, we have applied scalings to the corresponding vibrational transition dipole moments. This is a new approach implemented in \Trove\ which takes advantage of the $J=0$ representation of the basis set. Since the rovibrational line intensities are computed using  vibrational matrix elements of the electronically averaged dipole moment components $\bar{\mu}_{x}, \bar{\mu}_{y}$ and $\bar{\mu}_{z}$, modifying these vibrational moments by a scaling factor specific for a given band will propagate this scaling to all rotational lines within this band in a consistent manner. A band scaling factor was obtained as a geometric average of $n$ matched individual line intensities within each vibrational band:
\begin{equation}\label{eq:scale}
\bar{S} = \left[{\prod_{i=1}^{n}  \frac{I_i^{\rm HITRAN}}{I_i^{\rm TROVE}} }\right]^{\frac{1}{n}},
\end{equation}
which leads to a $\sqrt{\bar{S}}$ scaling factor on the dipole moment.
 To this end, we have correlated the HITRAN transition $T=$~296~K intensities to the corresponding intensities computed with TROVE using the methodology described above (after the band centre shifts).

\begin{figure}
	\includegraphics[width=0.97\linewidth]{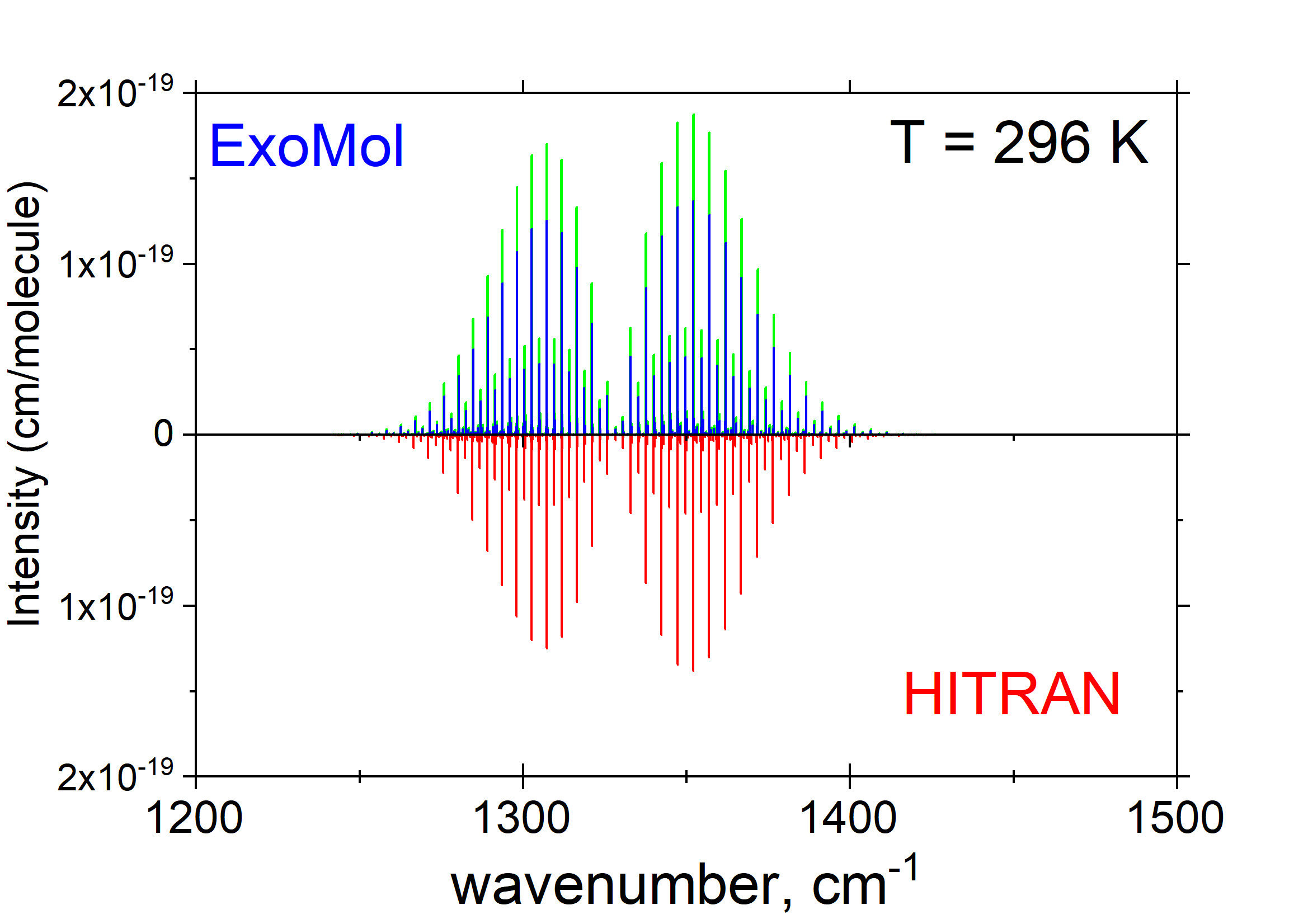}
	\caption{Comparison of aCeTY (this work) stick spectra, before (green) and after (blue) dipole moment scaling by a factor of 0.856,
	against HITRAN for the vibrational band (000110) $\Sigma_u^+$  of acetylene at $T = $~296~K.  }
	\label{fig:scaling:intens}
\end{figure}

Figure~\ref{fig:scaling:intens} shows an example of the dipole scaling procedure applied to the (000110) $\Sigma_u^+$ band. The unscaled TROVE (shown in green) intensities are $\sim$ 1.365 times stronger
than those from HITRAN, at 296~K. We can therefore apply a 0.856 ($=1/\sqrt{1.365}$) scaling factor to the $z$ dipole moment component of the corresponding vibrational matrix element $\langle000110|\mu_z|000000\rangle$, with the result shown in Figure~\ref{fig:scaling:intens} in blue. We have thus applied scaling factors to 216 bands, listed in the supplementary information to this work. Extracts are given in Tables~\ref{t:mu-scale-1}--\ref{t:mu-scale-3}.

\section{Results: the \protect{aCeTY} line list}\label{linelist}
	
\begin{figure}
			\includegraphics[width=\columnwidth]{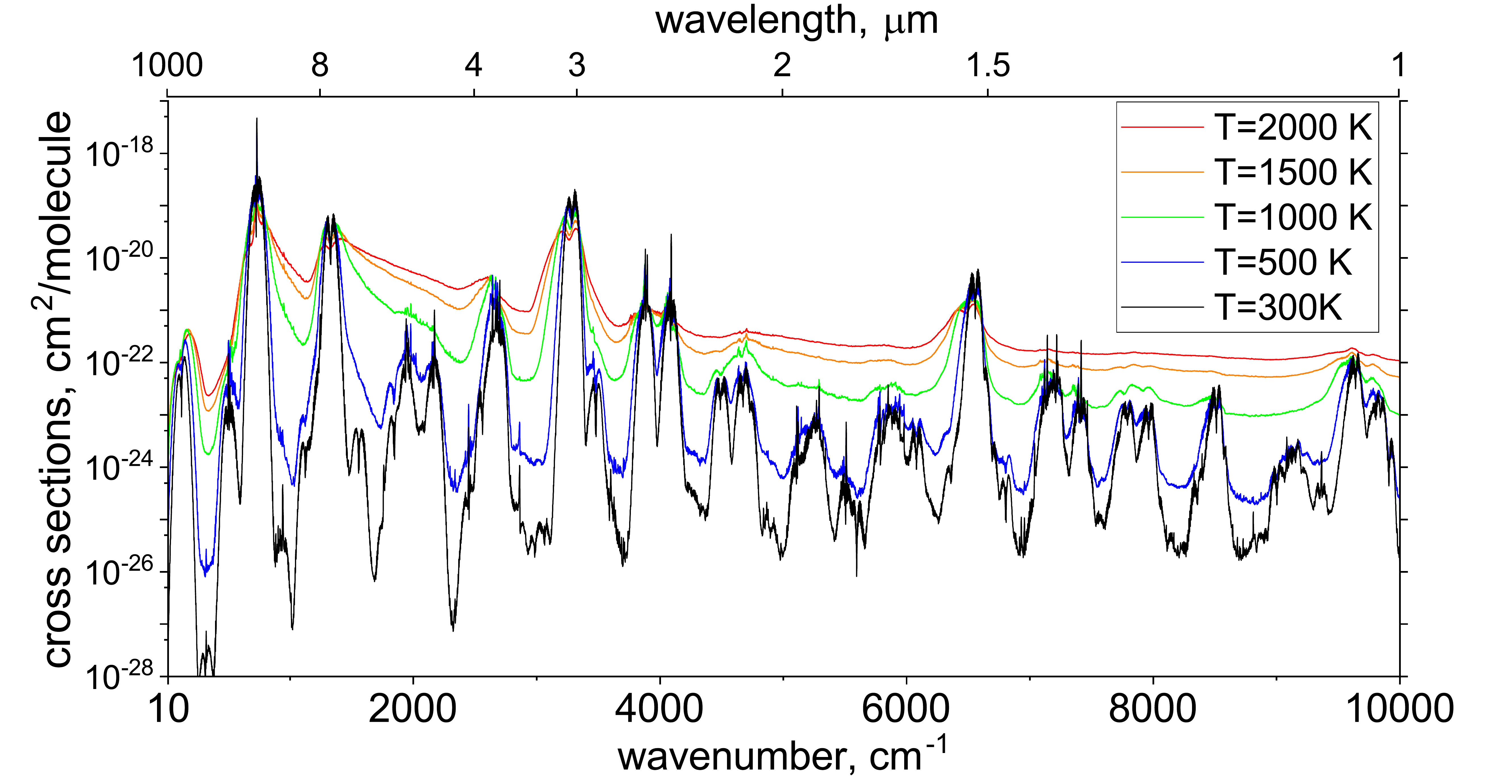}
			\caption{Variation of aCeTY line list spectra with temperature:
			low-resolution (1~cm$^{-1}$) cross-sections computed using
			ExoCross~\citep{jt708}. The spectrum becomes flatter with increasing temperature.}
			\label{fig:cross_secs}
		\end{figure}

The aCeTY line list has been computed using the variational calculations outlined above.
Figure~\ref{fig:cross_secs} illustrates the temperature-dependence of the acetylene spectra computed using the aCeTY line list, with cross-sections computed using ExoCross~\citep{jt708}
at a variety of temperatures between 296--2000~K. The cross-sections are calculated at a low-resolution of 1~\cm\ for demonstration purposes.

The \Trove\ assignment is based on the largest basis set contribution to the eigenfunction.
\Trove\ uses the  local mode quantum numbers (QNs) to assign vibrational state, collected in Table~\ref{t:QN}. This selection of the quantum numbers is based on the choice of the vibrational basis set in Eqs.~(\ref{e:class1_2}--\ref{e:class3_2}). There is no direct correlation between the local mode and normal mode assignment. For an approximation correlation, the following rules apply:
\begin{eqnarray}\label{e:normal:local}
  \upsilon_1 + \upsilon_3 &  = &  n_2 + n_3, \\
  \upsilon_2  &= &  n_1, \\
  \upsilon_4 + \upsilon_5   &=&  n_4+n_5+n_6+n_7.
\end{eqnarray}

\begin{figure}
	\includegraphics[width=\linewidth]{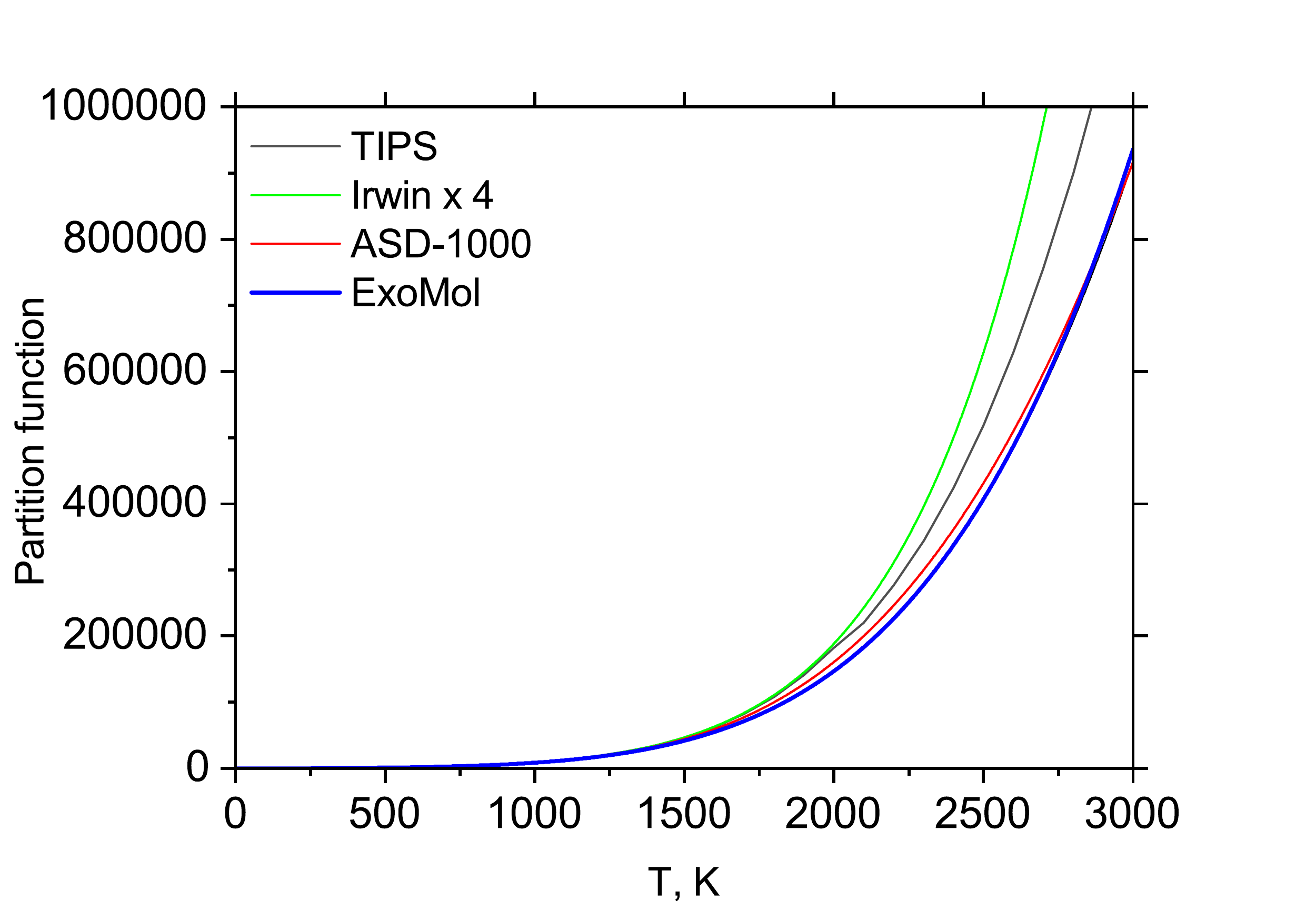}
	\caption{The \hcch\ partition function up to 3000~K; comparing the states from the aCeTY line list (this work) against that computed using TIPS~\citep{TIPS2017}, the coefficients of \citet{81Irwin}, and using energies extracted from ASD-1000 \citep{17LyPe.C2H2}.}
	\label{fig:comp_PF}
\end{figure}

The number density of a particular molecular state as a fraction of the total number density of the molecular species is given by the Boltzmann law.
The total internal partition function, $Q$, is a sum over all molecular states, weighting each by their probability of occupation at a given temperature, and therefore offers an indication of the completeness of a calculated line list at a particular temperature:
\begin{equation}\label{eq:pf}
Q = \sum_{i=1}^{N}{g_{ns}}^{(i)}(2 J_i+1) \exp\left({\frac{-c_2\tilde{E}_i}{T}}\right).
\end{equation}
Here, c$_2$=$\frac{hc}{k}$ is the second radiative constant, $\tilde{E}_i$ is
the energy term value of each $i$ molecular state (relative to the ground
ro-vibronic state), $T$ is the temperature, ${g_{ns}}^{(i)}$ is the nuclear
statistical weight of each $i$ molecular state, and the sum is over all
molecular states. The partition function can easily be computed from the ExoMol
states file~\citep{jt631} using ExoCross~\citep{jt708}. A comparison of the
partition function for \hcch\ computed using the aCeTY states file against the
partition function computed using TIPS~\citep{TIPS2017}, the coefficients by
\citet{81Irwin}
and the energies extracted from the ASD-1000 database of \citet{17LyPe.C2H2}  is
given in Figure~\ref{fig:comp_PF}.  All these partition functions, with the
exception of the one due to \citep{81Irwin}, use
the ``physicists'' convention which weights ortho and para states of \hcch\ 3 and 1, respectively,
as opposed to the weighting of 0.75 and 0.25 often adopted by astronomers.

\begin{figure}
	\includegraphics[width=\linewidth]{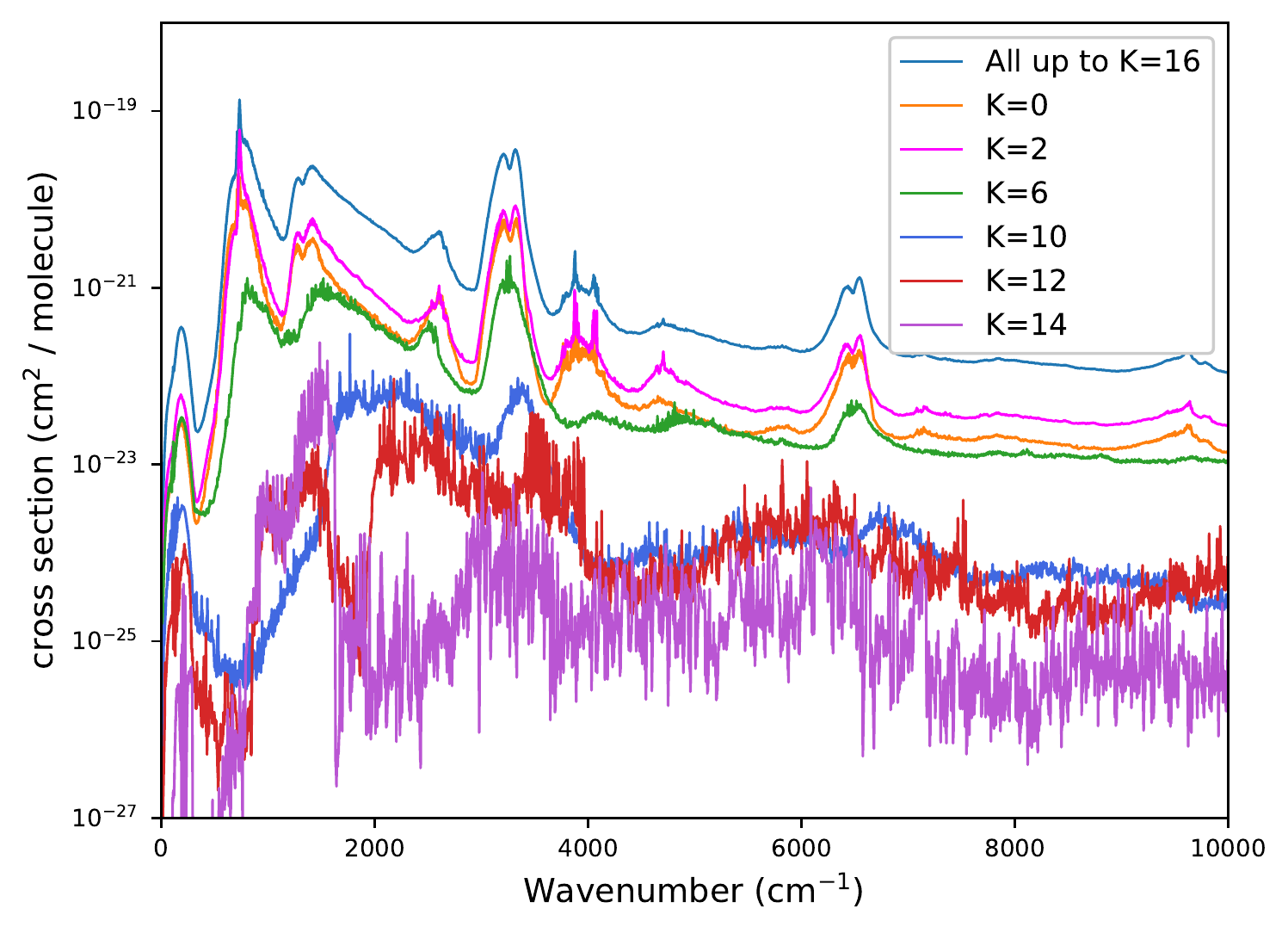}
	\caption{Contribution from transitions with upper states of different $K'$ to the aCeTY opacity at $T$~=~2000~K. Note that the $K'$~=~16 transitions do not appear on this scale.  }
	\label{fig:C2H2_2000K_K10}
\end{figure}

\begin{figure}
	\includegraphics[width=\linewidth]{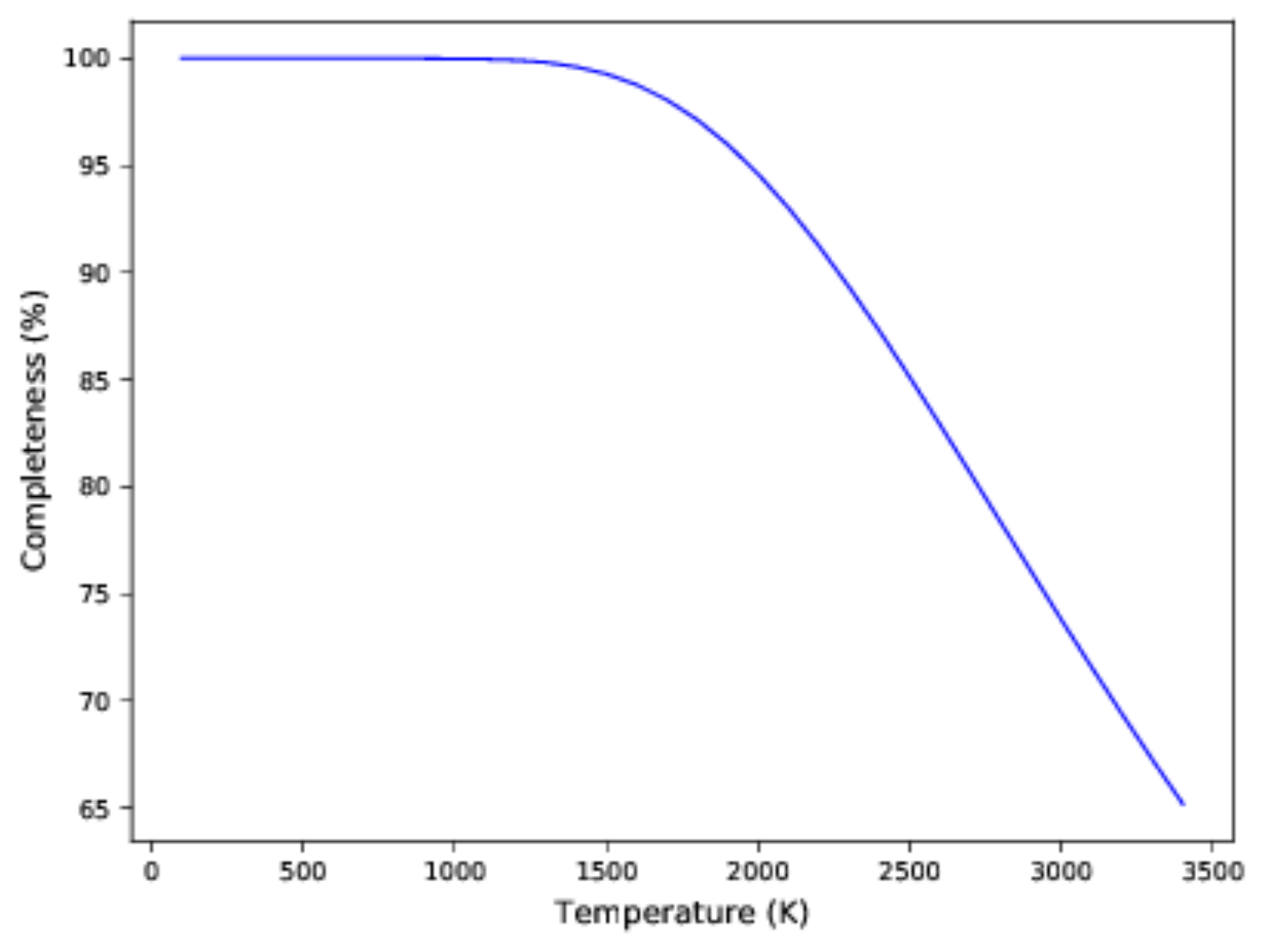}
	\caption{The completeness of the aCeTY line list as a function of temperature, up to 3400~K.  }
	\label{t:temp_dep}
\end{figure}

The temperature at which a line list is complete up to is generally dictated by
the range of energy levels included in a line list production. For the aCeTY
line list we use a maximum lower energy of 12~000~cm$^{-1}$, and a maximum upper
energy of 22~000~cm$^{-1}$, which gives a line list that is complete up to
10~000~cm$^{-1}$ (i.e. $\lambda > 1$~$\mu$m), i.e. the maximum upper energy is
10~000~cm$^{-1}$ above the maximum lower energy level value. The  completeness as a
function of temperature
of such a line list can be estimated by calculating the partition
function up to the lower energy level cut-off as a percentage of the total
partition function which includes all states involved in a line list
calculation.
Figure~\ref{t:temp_dep} gives these values at a variety of temperatures.
A line list is generally considered to be ``complete'' if the ratio of the
partition function of the lower energy states to the partition function of all
energy states involved in a line list calculation is $>90\%$. It can be seen
from Figure~\ref{t:temp_dep} that the aCeTY line list is therefore estimated,
using this metric, to be complete up to around 2200~K.
The other factor which could have an impact on the completeness of the line list is the value used for $K_{\rm max}=L_{\rm max}$ in Eq.~(\ref{e:lmax_kmax}). As discussed, however, we do not expect the states below 22~000~\cm\ which are being used for the line list to have high values of $L_{\rm max}$; the bending states which correspond to a high value of $L$ are expected at very high energies, and therefore temperatures.
Figure~\ref{fig:C2H2_2000K_K10} shows the contributions from transitions with upper states with different values of $K$ ($=L$) to a line list computed up to $K_{\rm max}=L_{\rm
max}$~=~16. It can be seen that states with higher values of $K$ contribute a
vanishingly small amount to the overall opacity. We therefore do not expect
increasing the value of $K_{\rm max}=L_{\rm max}$ in a calculation to have a significant
effect on the opacity  of a acetylene even at high temperatures.

A complete description of the ExoMol data structure along with examples was reported by \citet{jt631}.
The ExoMol .states file contains all computed ro-vibrational energies (in cm$^{-1}$) relative to the ground state. Each energy level
is assigned a unique state ID with symmetry and quantum number labelling; an extract for \hcch\ is shown in Table \ref{tab.states}. The .trans files,
which are split into frequency windows for ease of use, contain all computed transitions with upper
and lower state ID labels, and Einstein A coefficients. An example from a .trans file for the aCeTY line list is given in
Table \ref{tab.tran}.

\renewcommand{\tabcolsep}{0.15cm}
\begin{table*}
\caption{Extract from the  .states file for the aCeTY line list. The theoretical energies were replaced with MARVEL where available (see text).}
\label{tab.states}
\begin{center}
\begin{tabular}{rrrrrrrrrrrrrrrrrr}
\hline
$N$ & \multicolumn{1}{c}{$\tilde{E}$} & $g_{\text{tot}}$ &  $J$ & Unc.  & $\Gamma_{\text{tot}}$ & $n_1$ & $n_2$ & $n_3$ & $n_4$ & $n_5$ & $n_6$ & $n_7$ &
$\Gamma_{\text{vib}}$ & $K$ & $\tau_{\text{rot}}$ & $\Gamma_{\text{rot}}$ &$\tilde{E}^{\rm TROVE}$ \\
\hline
793042	&	14.119512	&	21	&	3	&	0.000041	&	A2g	&	0	&	0	&	0	&	0	&	0	&	0	&	0	&	A1g	&	0	&	1	&	A2g	&	14.117789	\\
793043	&	625.7974	&	21	&	3	&	0.000204	&	A2g	&	0	&	0	&	0	&	0	&	0	&	0	&	1	&	E1g	&	1	&	1	&	E1g	&	625.793876	\\
793044	&	1242.947091	&	21	&	3	&	0.000434	&	A2g	&	0	&	0	&	0	&	0	&	1	&	0	&	1	&	E2g	&	2	&	0	&	E2g	&	1242.933977	\\
793045	&	1244.544497	&	21	&	3	&	0.000368	&	A2g	&	0	&	0	&	0	&	1	&	0	&	1	&	0	&	A1g	&	0	&	1	&	A2g	&	1244.546193	\\
793046	&	1463.286132	&	21	&	3	&	0.000266	&	A2g	&	0	&	0	&	0	&	0	&	0	&	0	&	2	&	A1g	&	0	&	1	&	A2g	&	1463.285595	\\
793047	&	1472.462847	&	21	&	3	&	0.000532	&	A2g	&	0	&	0	&	0	&	0	&	1	&	0	&	1	&	E2g	&	2	&	0	&	E2g	&	1472.455065	\\
793048	&	1865.740794	&	21	&	3	&	1	&	A2g	&	0	&	0	&	0	&	1	&	0	&	1	&	1	&	E3g	&	3	&	1	&	E3g	&	-1	\\
793049	&	1868.702371	&	21	&	3	&	0.001250	&	A2g	&	0	&	0	&	0	&	0	&	2	&	0	&	1	&	E1g	&	1	&	1	&	E1g	&	1868.70286	\\
793050	&	1988.361458	&	21	&	3	&	0.001649	&	A2g	&	1	&	0	&	0	&	0	&	0	&	0	&	0	&	A1g	&	0	&	1	&	A2g	&	1988.353787	\\
793051	&	2062.011051	&	21	&	3	&	0.001087	&	A2g	&	0	&	0	&	0	&	0	&	3	&	0	&	0	&	E1g	&	1	&	1	&	E1g	&	2062.010327	\\
793052	&	2080.003537	&	21	&	3	&	0.001923	&	A2g	&	0	&	0	&	0	&	2	&	0	&	0	&	1	&	E1g	&	1	&	1	&	E1g	&	2080.003632	\\
793053	&	2088.260006	&	21	&	3	&	1	&	A2g	&	0	&	0	&	0	&	2	&	1	&	0	&	0	&	E3g	&	3	&	1	&	E3g	&	-1	\\
793054	&	2500.471967	&	21	&	3	&	2	&	A2g	&	0	&	0	&	0	&	0	&	2	&	0	&	2	&	E2g	&	2	&	0	&	E2g	&	-1	\\
793055	&	2502.756348	&	21	&	3	&	2	&	A2g	&	0	&	0	&	0	&	2	&	0	&	2	&	0	&	A1g	&	0	&	1	&	A2g	&	-1	\\
793056	&	2587.563753	&	21	&	3	&	0.001414	&	A2g	&	1	&	0	&	0	&	0	&	0	&	0	&	1	&	E1g	&	1	&	1	&	E1g	&	2587.571805	\\
793057	&	2662.227679	&	21	&	3	&	0.004000	&	A2g	&	0	&	0	&	0	&	0	&	2	&	0	&	2	&	A1g	&	0	&	1	&	A2g	&	2662.229753	\\
793058	&	2675.549505	&	21	&	3	&	2	&	A2g	&	0	&	0	&	0	&	2	&	0	&	2	&	0	&	E2g	&	2	&	0	&	E2g	&	-1	\\
793059	&	2699.065141	&	21	&	3	&	2	&	A2g	&	0	&	0	&	0	&	2	&	0	&	0	&	2	&	A1g	&	0	&	1	&	A2g	&	-1	\\
793060	&	2703.164786	&	21	&	3	&	2	&	A2g	&	0	&	0	&	0	&	2	&	1	&	0	&	1	&	E2g	&	2	&	0	&	E2g	&	-1	\\
793061	&	2894.449628	&	21	&	3	&	0.000970	&	A2g	&	0	&	0	&	0	&	4	&	0	&	0	&	0	&	A1g	&	0	&	1	&	A2g	&	2894.449274	\\
\hline
\end{tabular}
\end{center}

\mbox{}\\

{\flushleft
$N$ : State ID; \\
$\tilde{E}$: Term value (in cm$^{-1}$);\\
$g_{\text{tot}}$: Total degeneracy; \\
$J$: Total angular momentum;\\
$\Gamma_{\text{tot}}$: Total symmetry in \Dh{\infty}(M)  \\
$n_1$-$n_{7}$: \trove\ vibrational quantum numbers (QN) (see Eq.~\eqref{e:normal:local});\\
$\Gamma_{\text{vib}}$: Symmetry of vibrational component of state in D$_{\infty h}$(M); \\
$K$: Projection of $J$ on molecule-fixed $z$-axis ($K=L$); \\
$\tau_{\text{rot}}$: Rotational parity (0 or 1); \\
$\Gamma_{\text{rot}}$: Symmetry of rotational component of state in D$_{\infty h}$(M);\\
$\tilde{E}^{\rm TROVE}$: \Trove\ term value, if replaced with MARVEL (in cm$^{-1}$);\\
Unc.: Uncertainty (\cm).\\

}
\end{table*}

\renewcommand{\arraystretch}{1.1}
\renewcommand{\tabcolsep}{0.5cm}
\begin{table}
\caption{Extract from a .trans file for the aCeTY line list.  }
\begin{tabular}{r r r }
\hline
\multicolumn{1}{c}{$f$} & $i$  & \multicolumn{1}{c}{$A_{fi}$}  \\
\hline
      &       &             \\
       56645     &      1    &  2.19617648E-05   \\
       56554     &      1    &  4.23999064E-02   \\
       56923     &      1    &  3.00213998E-03   \\
       56736     &      1    &  8.28753699E-09   \\
       56839     &      1    &  8.60646693E-07   \\
       56646    &       1   &   1.25543720E-05   \\
       56555    &       1   &   8.33151532E-04   \\
       56924    &       1   &   2.58105436E-06   \\

\hline
\end{tabular}
\label{tab.tran}
\mbox{}\\

$f$: Upper  state ID; \\
$i$: Lower  state ID; \\
$A_{fi}$: Einstein A coefficient (in s$^{-1}$).


\end{table}

\section{Comparison to Other Data}\label{compare}
		
\begin{figure}
	\includegraphics[width=\linewidth]{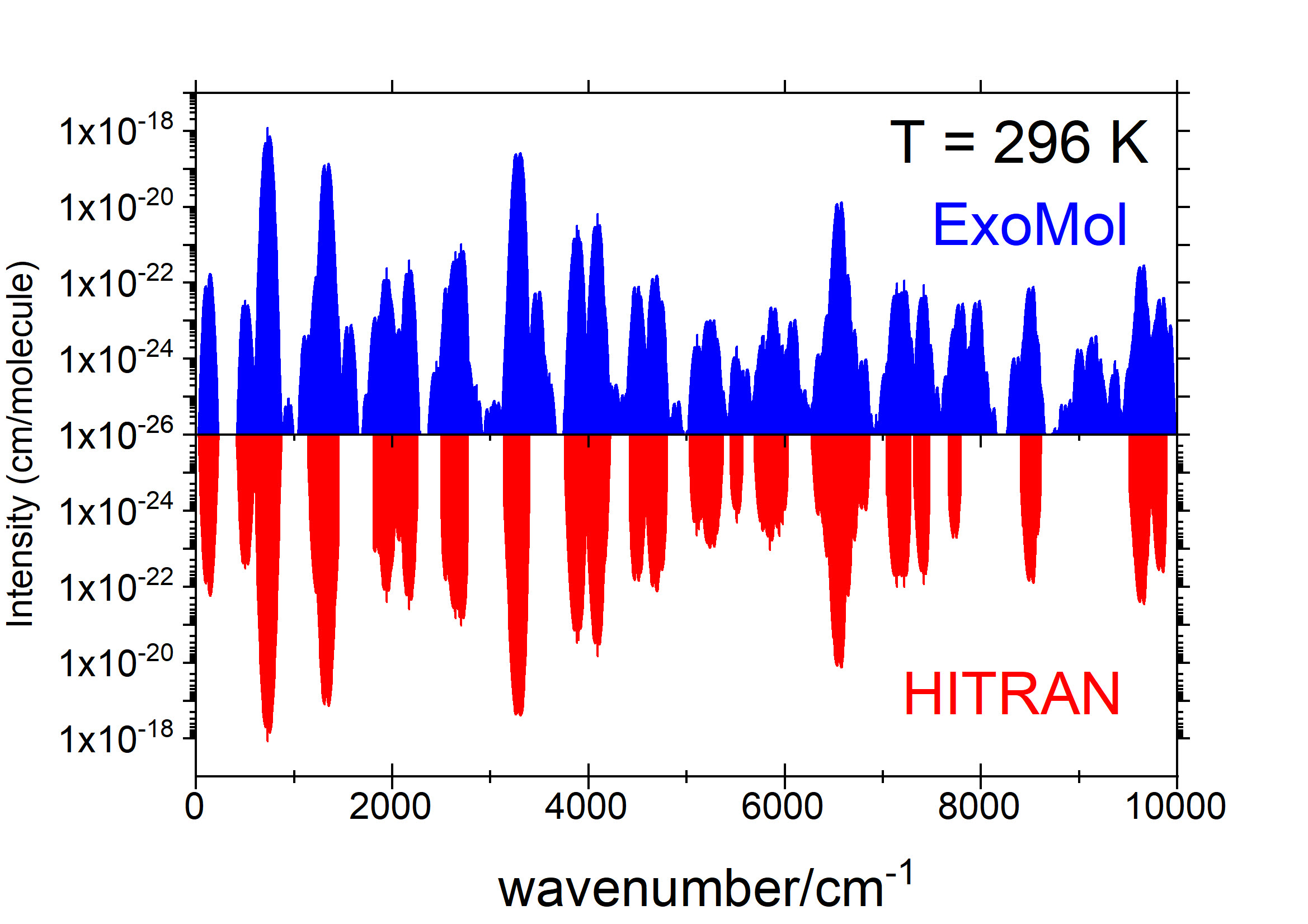}
	\caption{Comparison of the aCeTY stick spectrum with the HITRAN data for the range up to 10~000~\cm\ at 296~K.  }
	\label{fig:HITRAN}
\end{figure}

\begin{figure*}
	\includegraphics[width=0.46\linewidth]{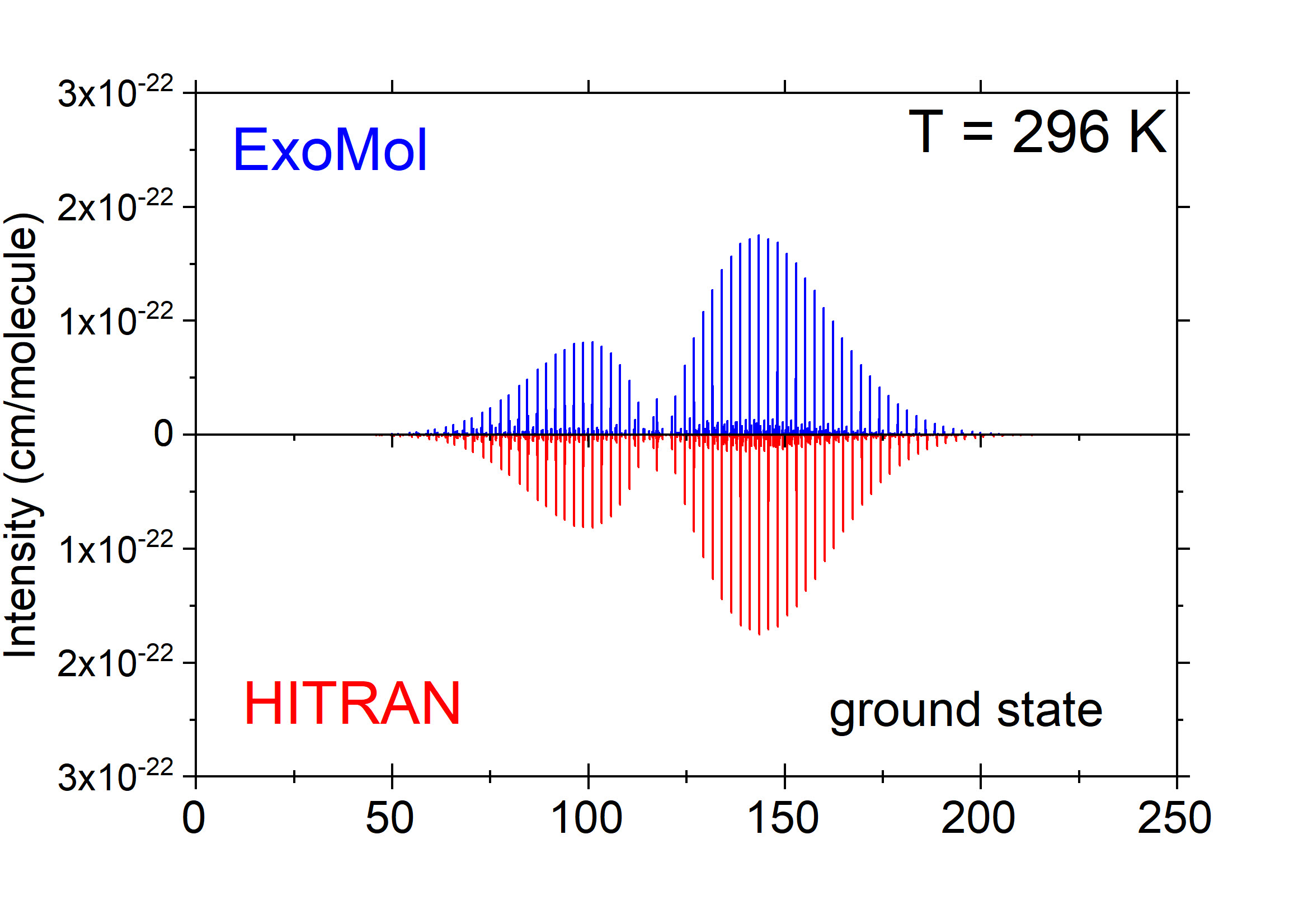}
	\includegraphics[width=0.46\linewidth]{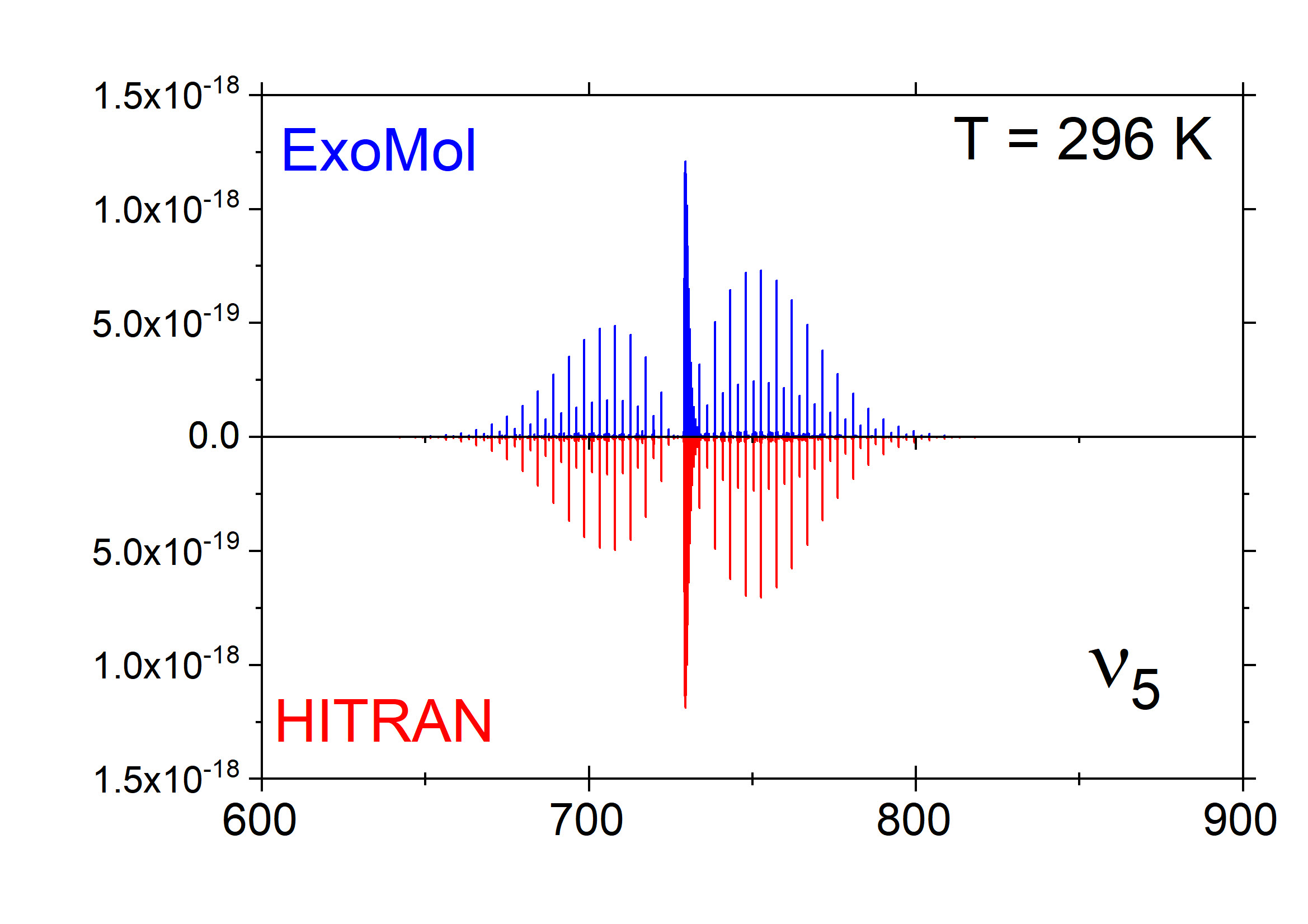}
	\includegraphics[width=0.46\linewidth]{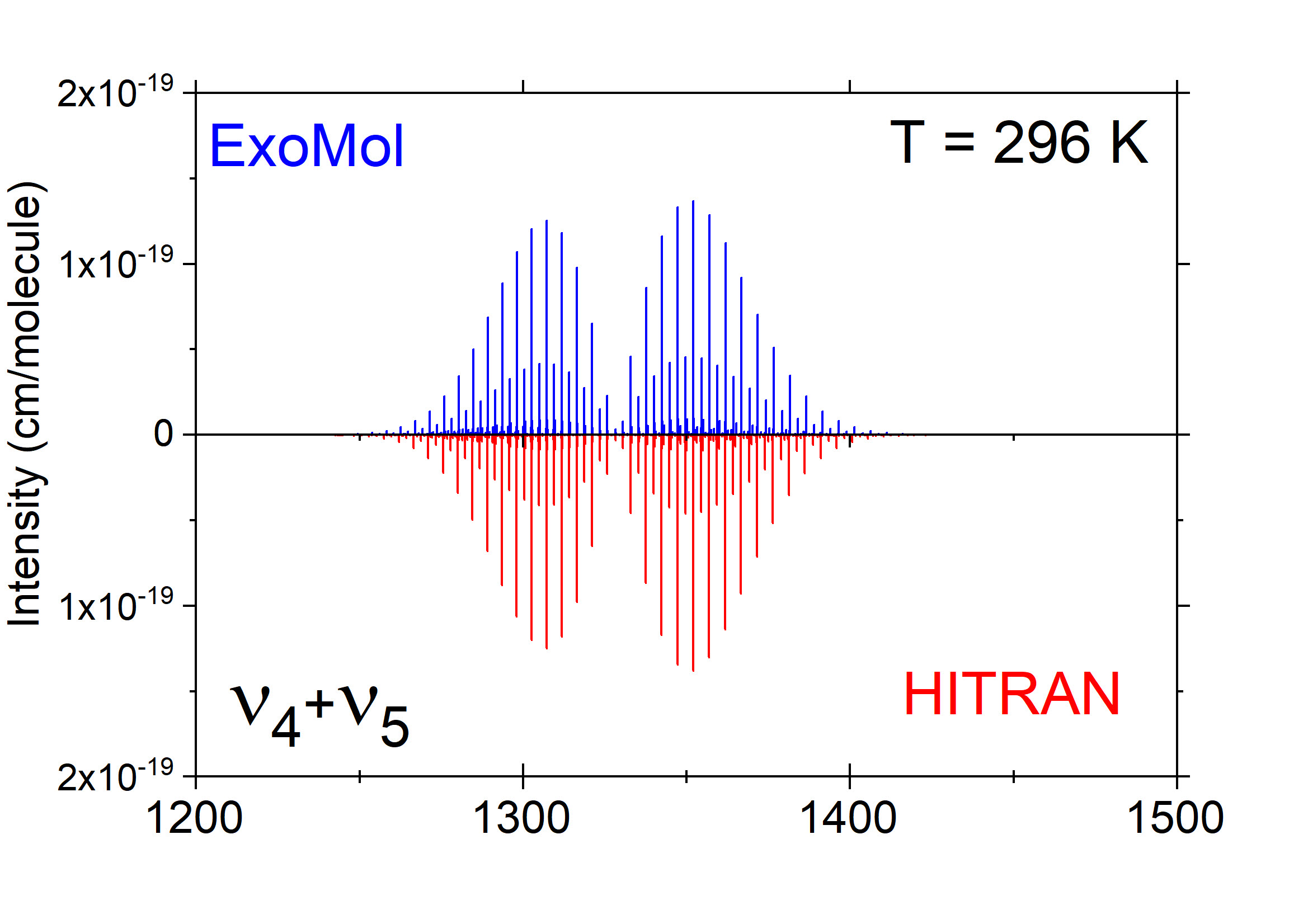}
	\includegraphics[width=0.46\linewidth]{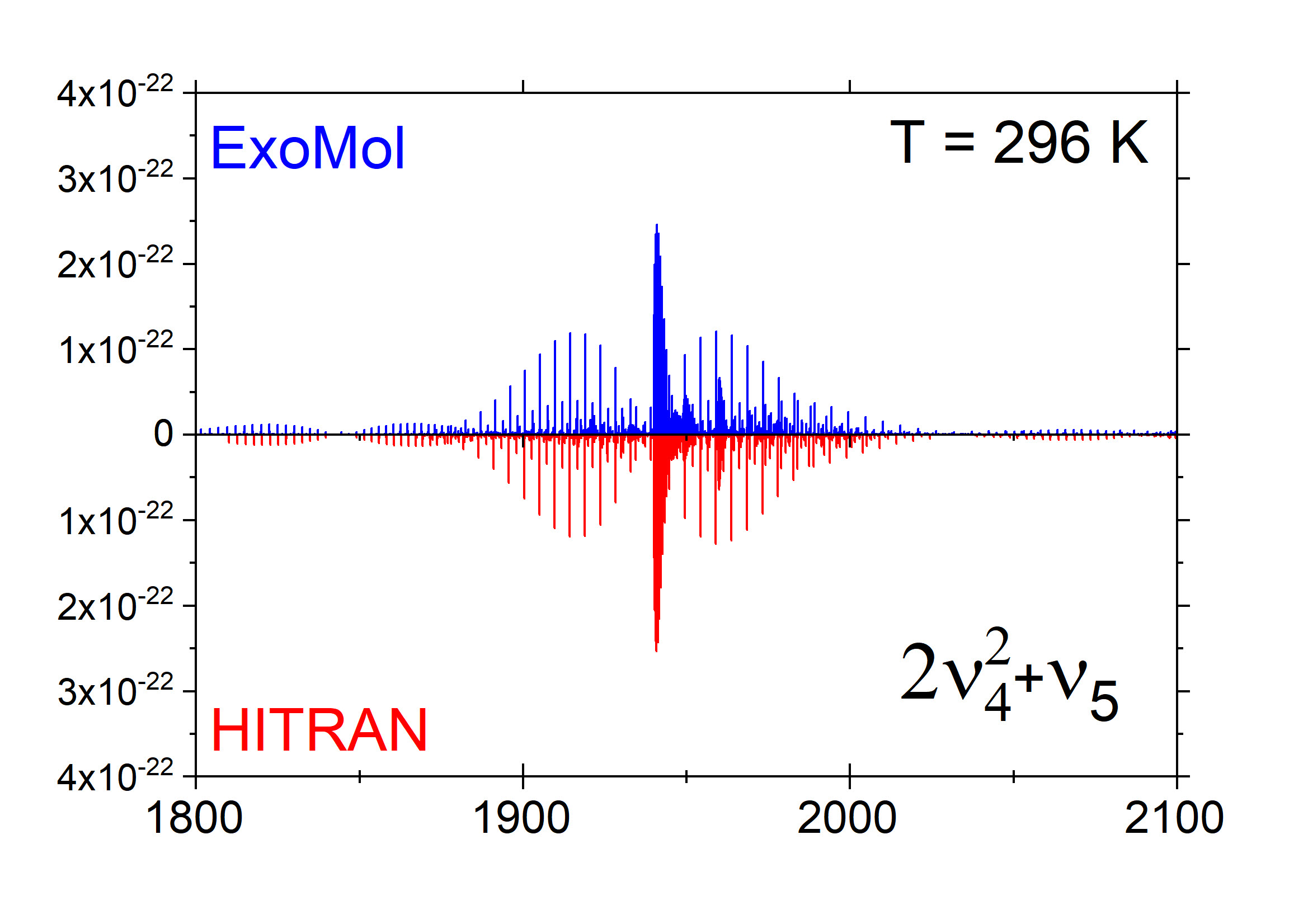}
	\includegraphics[width=0.46\linewidth]{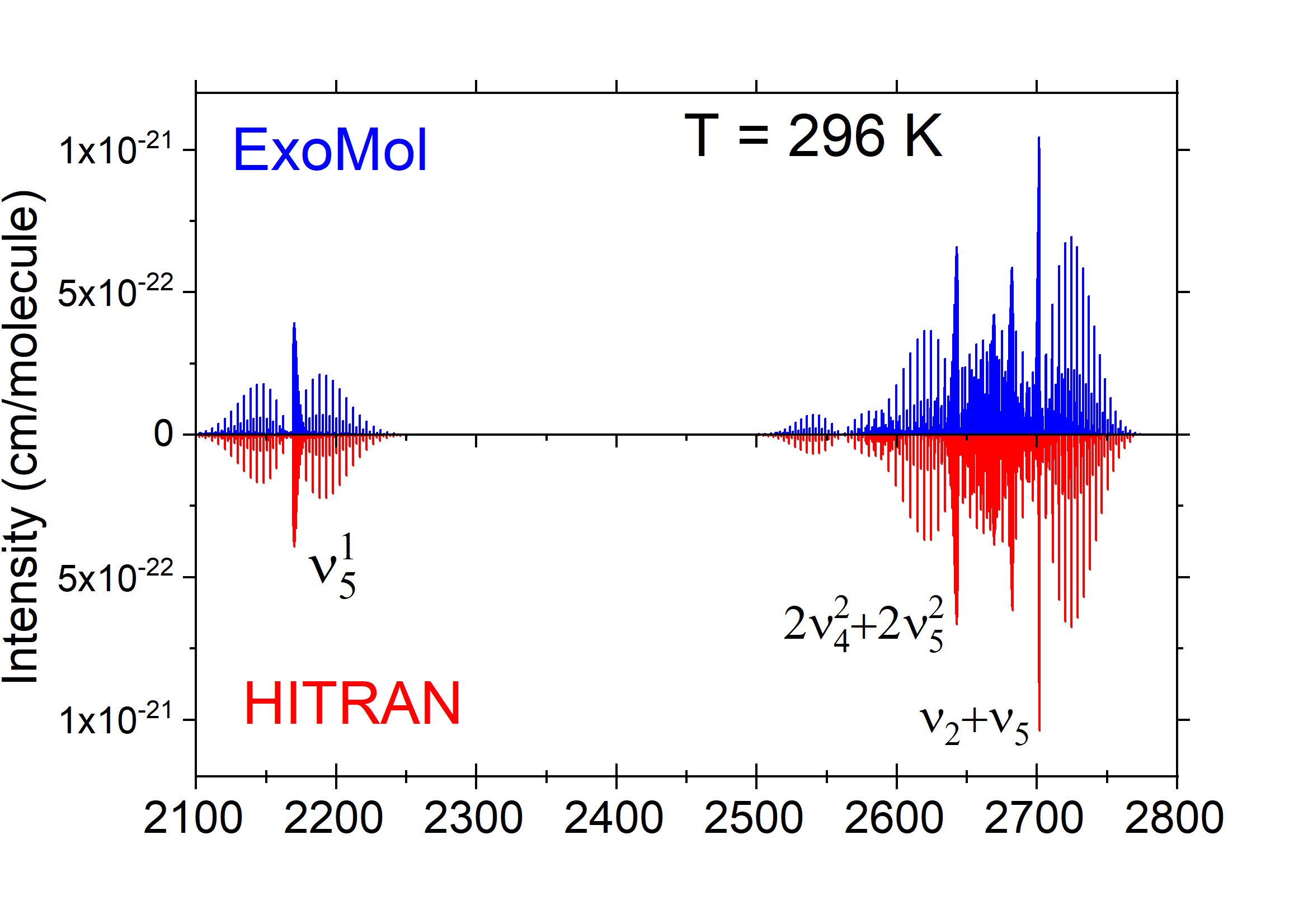}
	\includegraphics[width=0.46\linewidth]{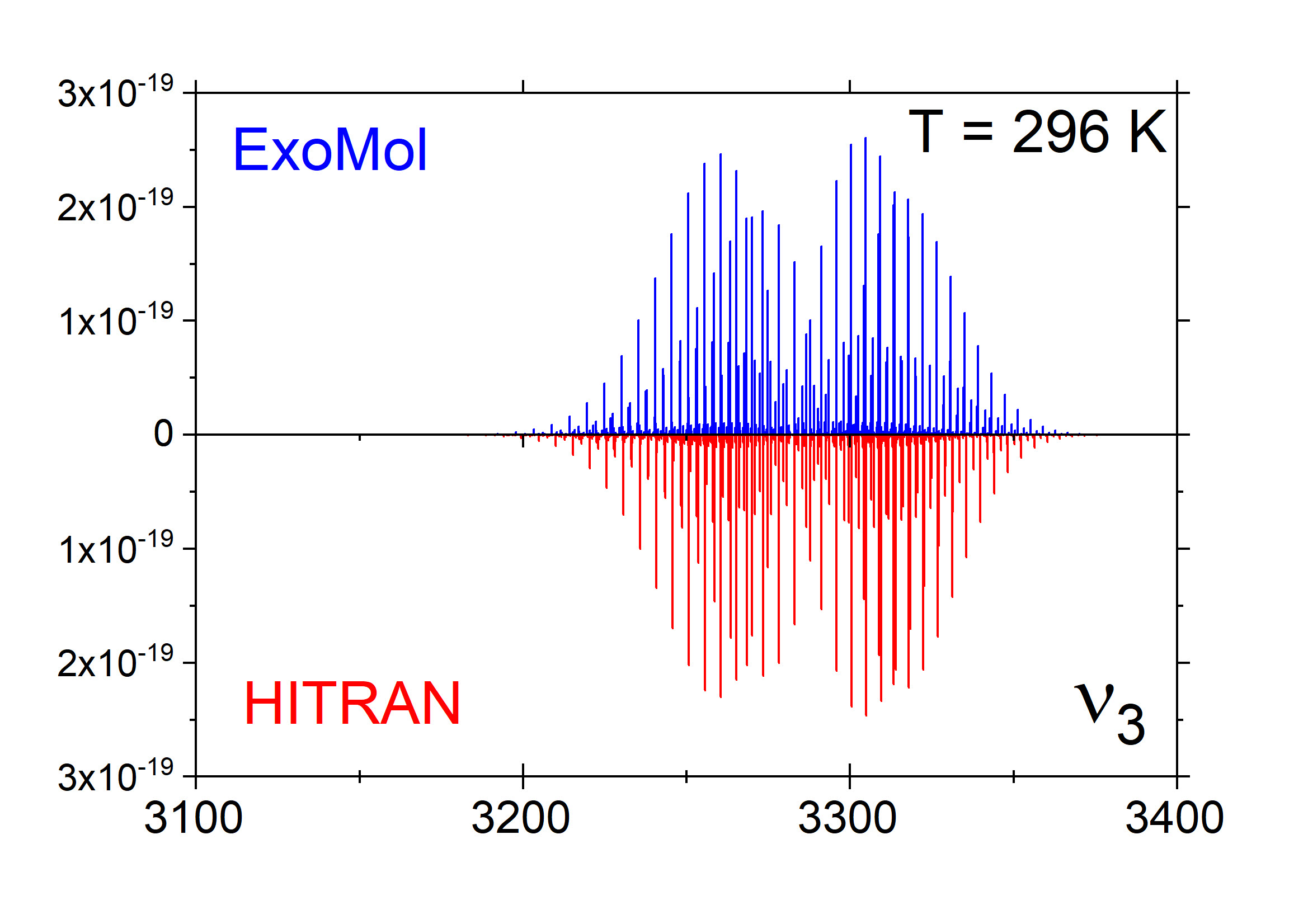}
	\includegraphics[width=0.46\linewidth]{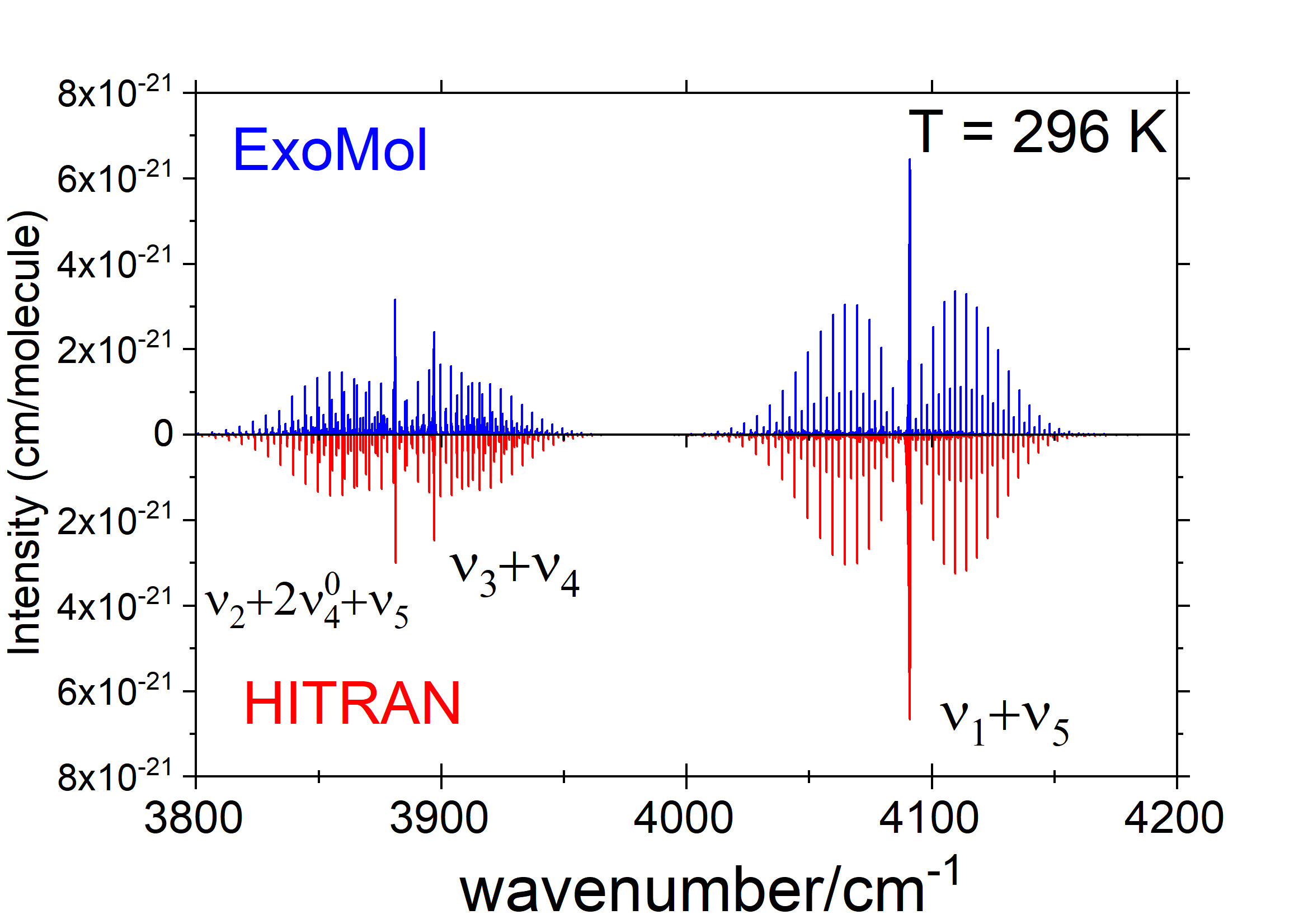}
	\includegraphics[width=0.46\linewidth]{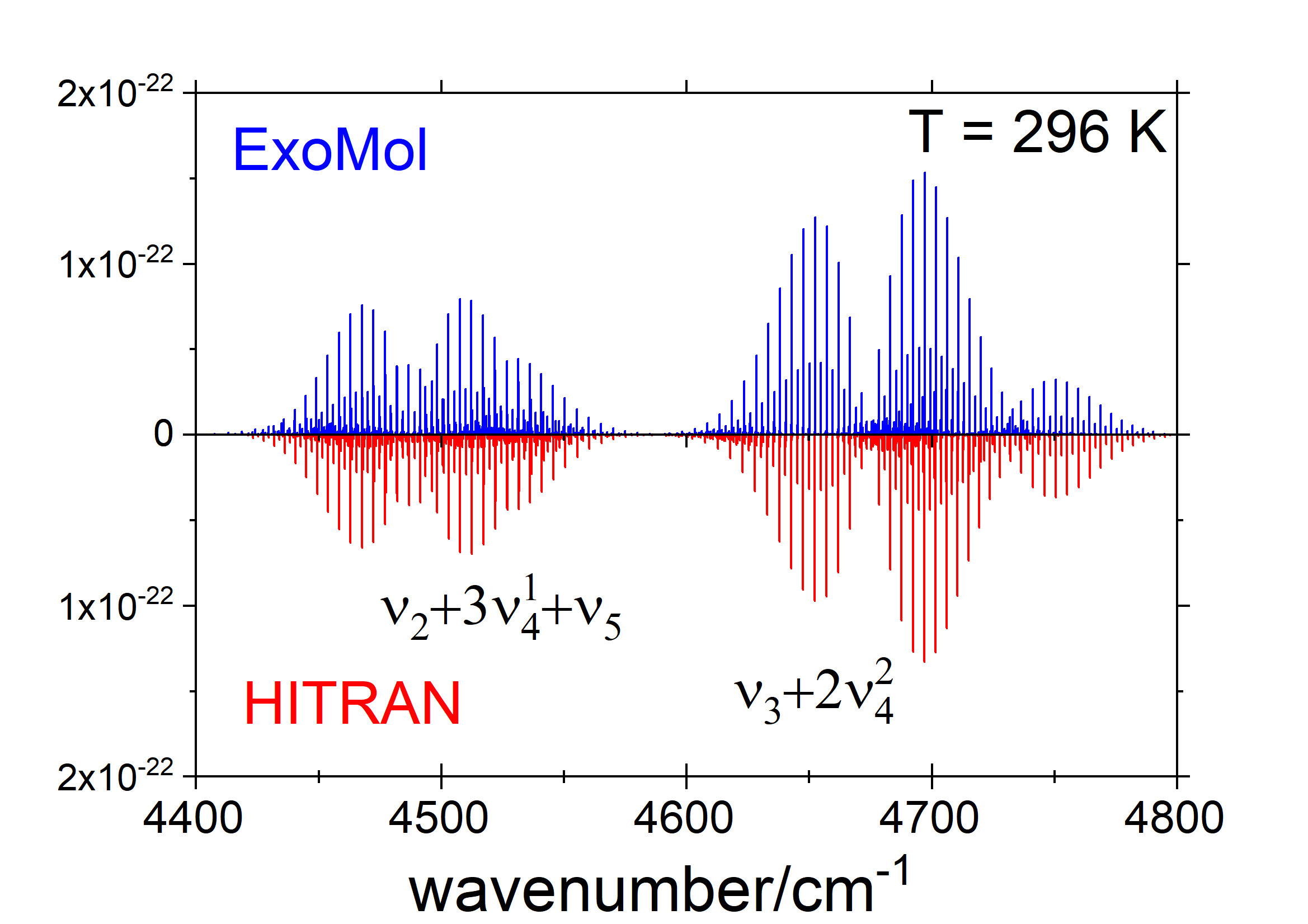}
	\caption{Comparison of the aCeTY stick spectra (with scaling of the dipole moment applied) against HITRAN for different vibrational bands of acetylene at 296~K (in the range 0-5000~\cm). The vibrational assignment of  the strongest bands is shown.}
	\label{fig:HITRAN:all}
\end{figure*}

\begin{figure*}
	\includegraphics[width=0.48\linewidth]{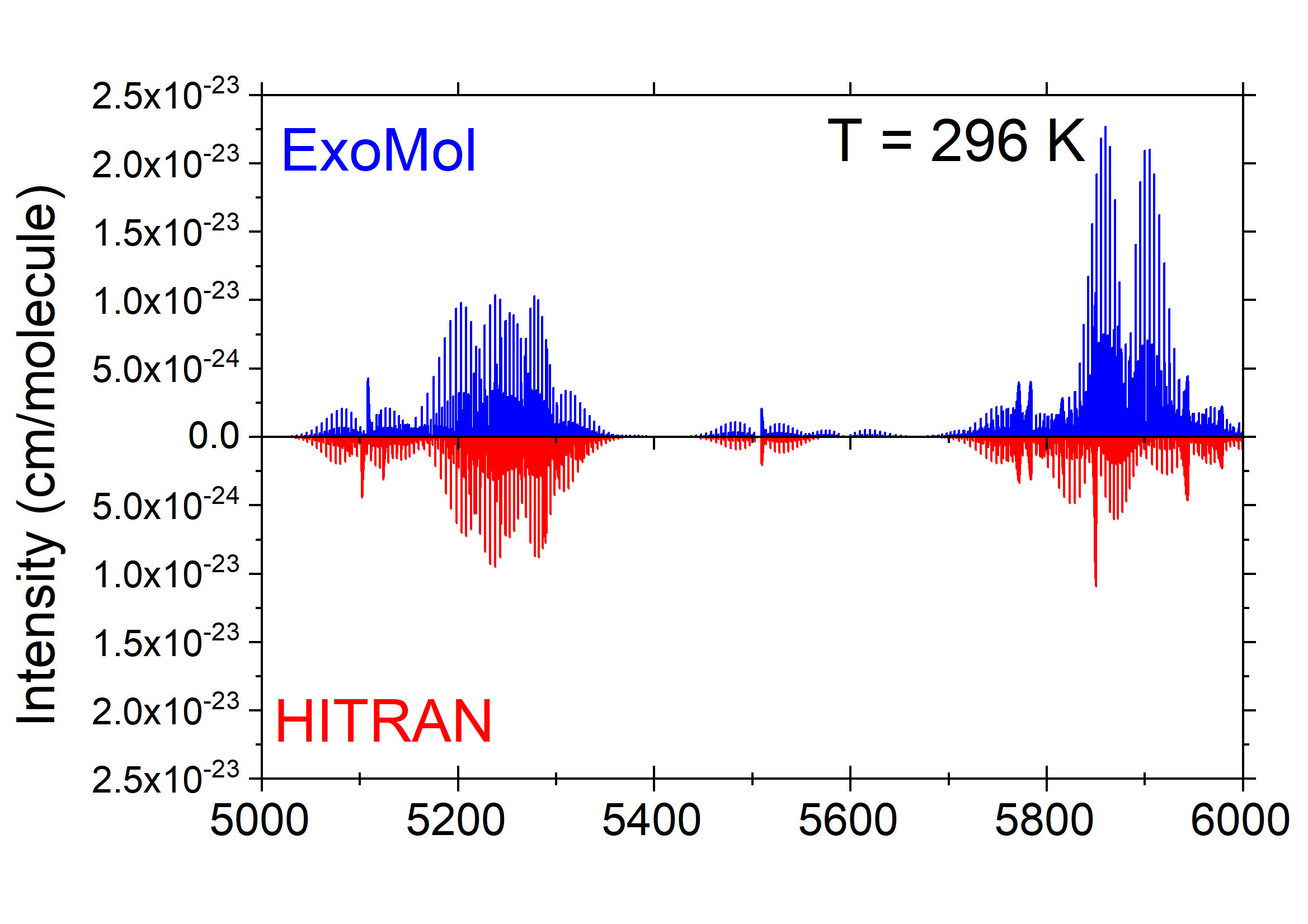}
	\includegraphics[width=0.48\linewidth]{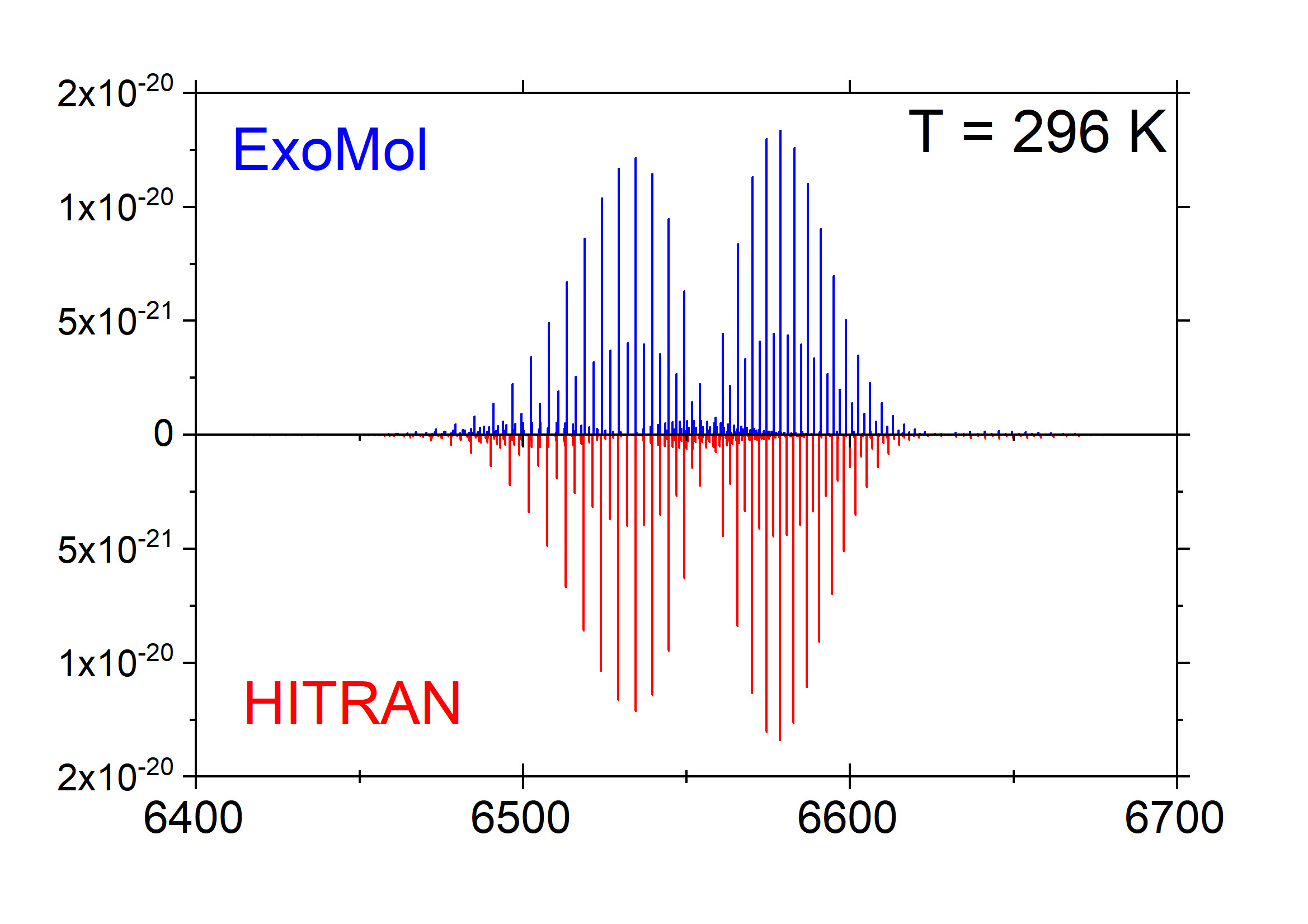}
	\includegraphics[width=0.48\linewidth]{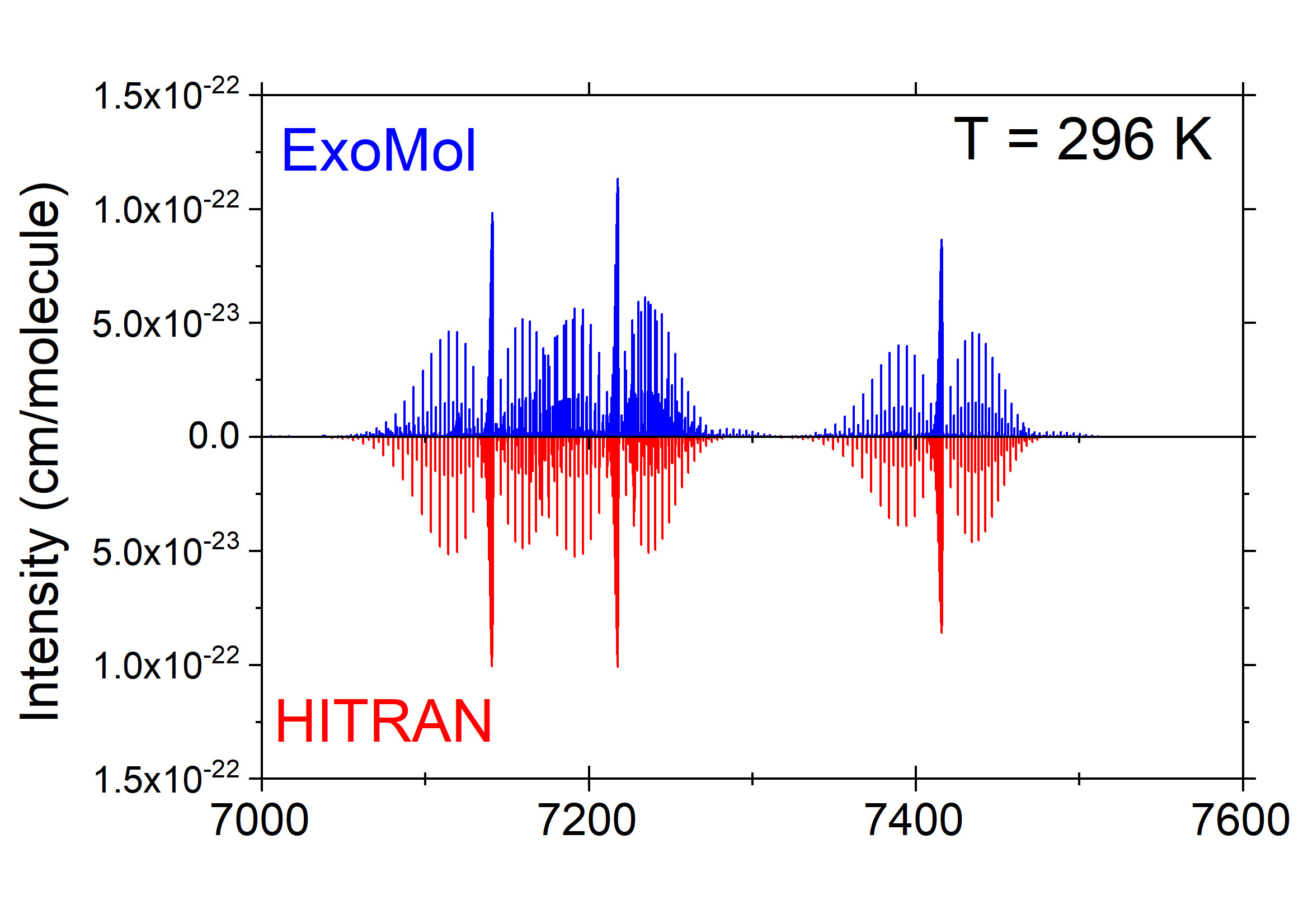}
	\includegraphics[width=0.48\linewidth]{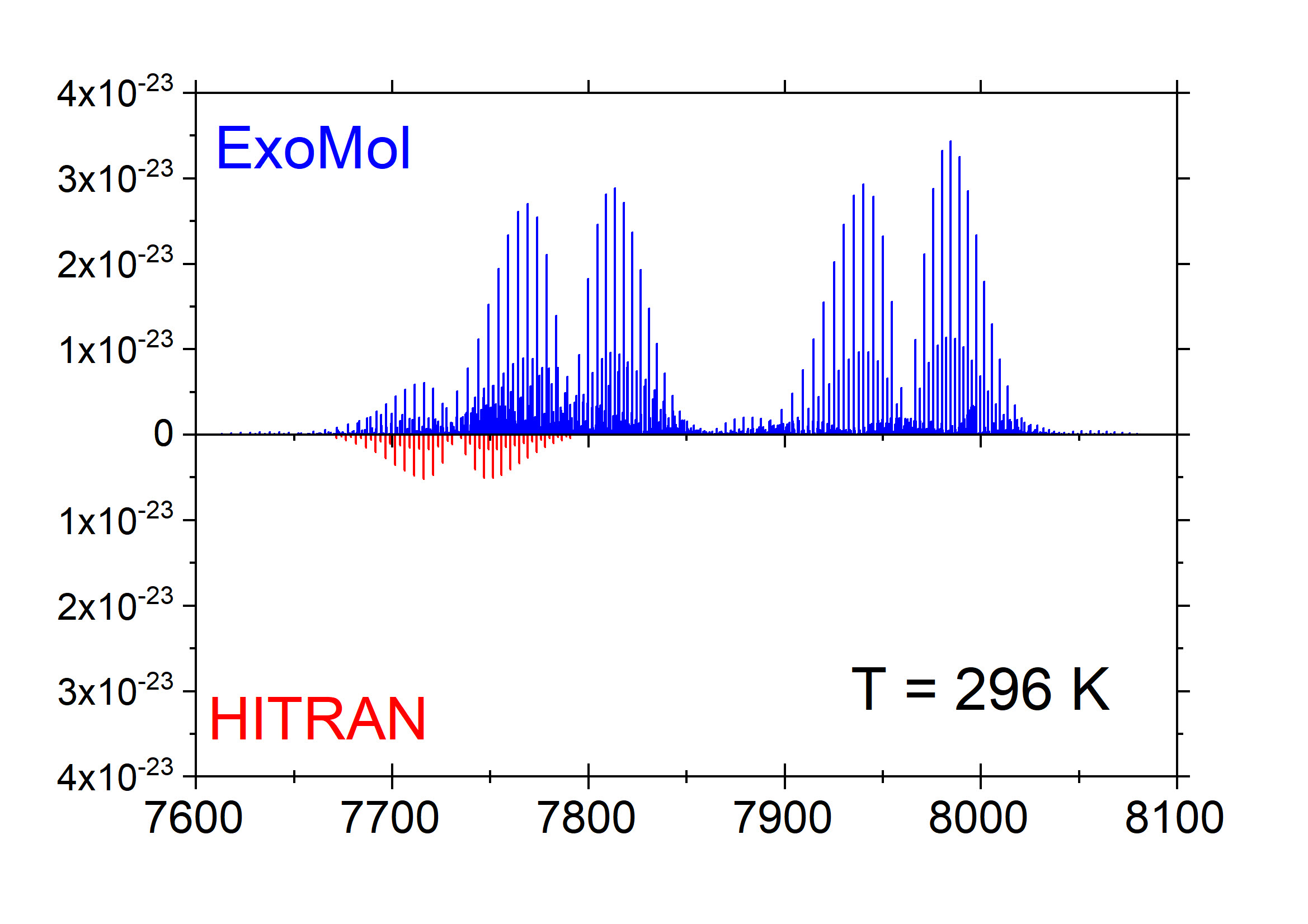}
	\includegraphics[width=0.48\linewidth]{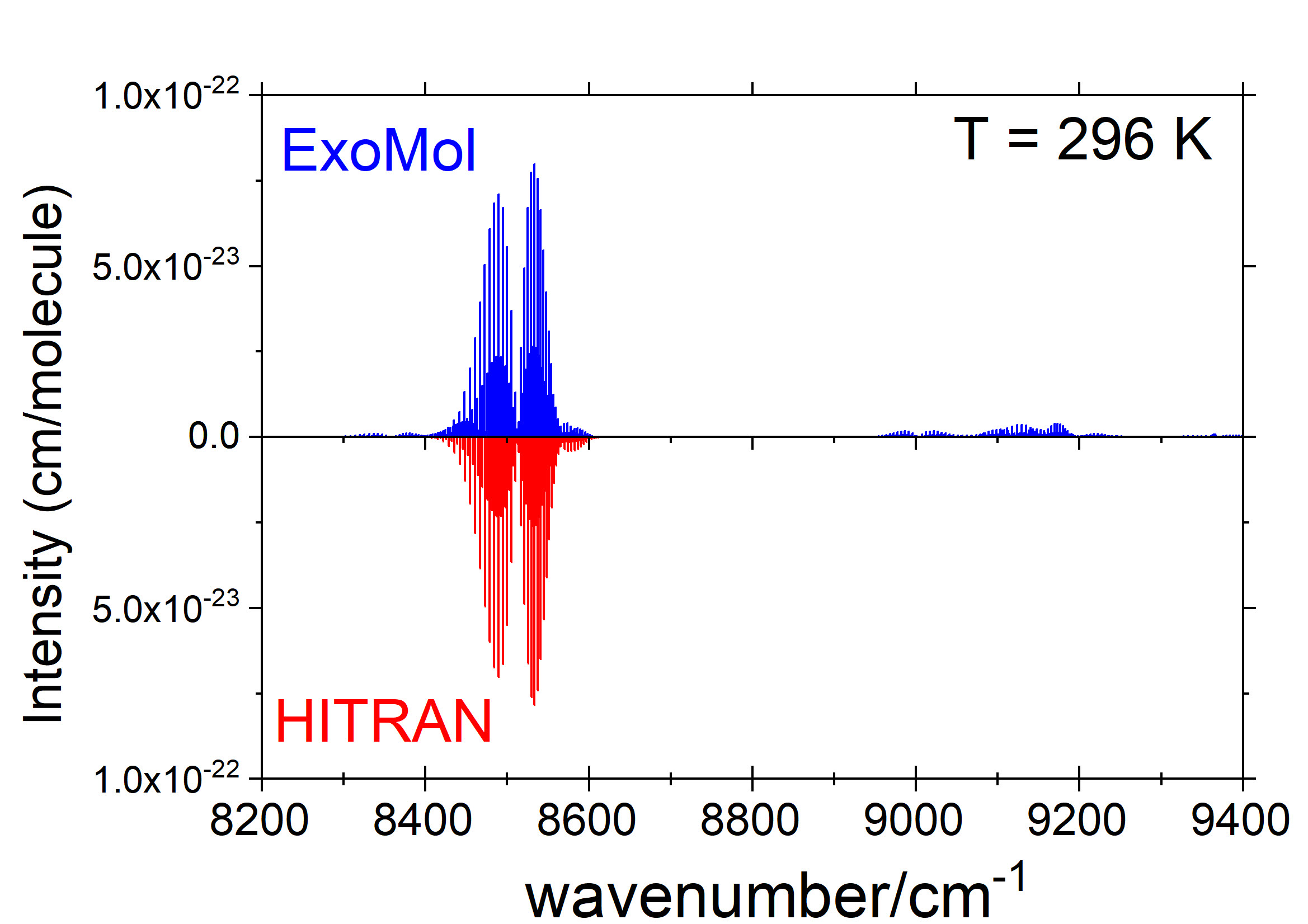}
	\includegraphics[width=0.48\linewidth]{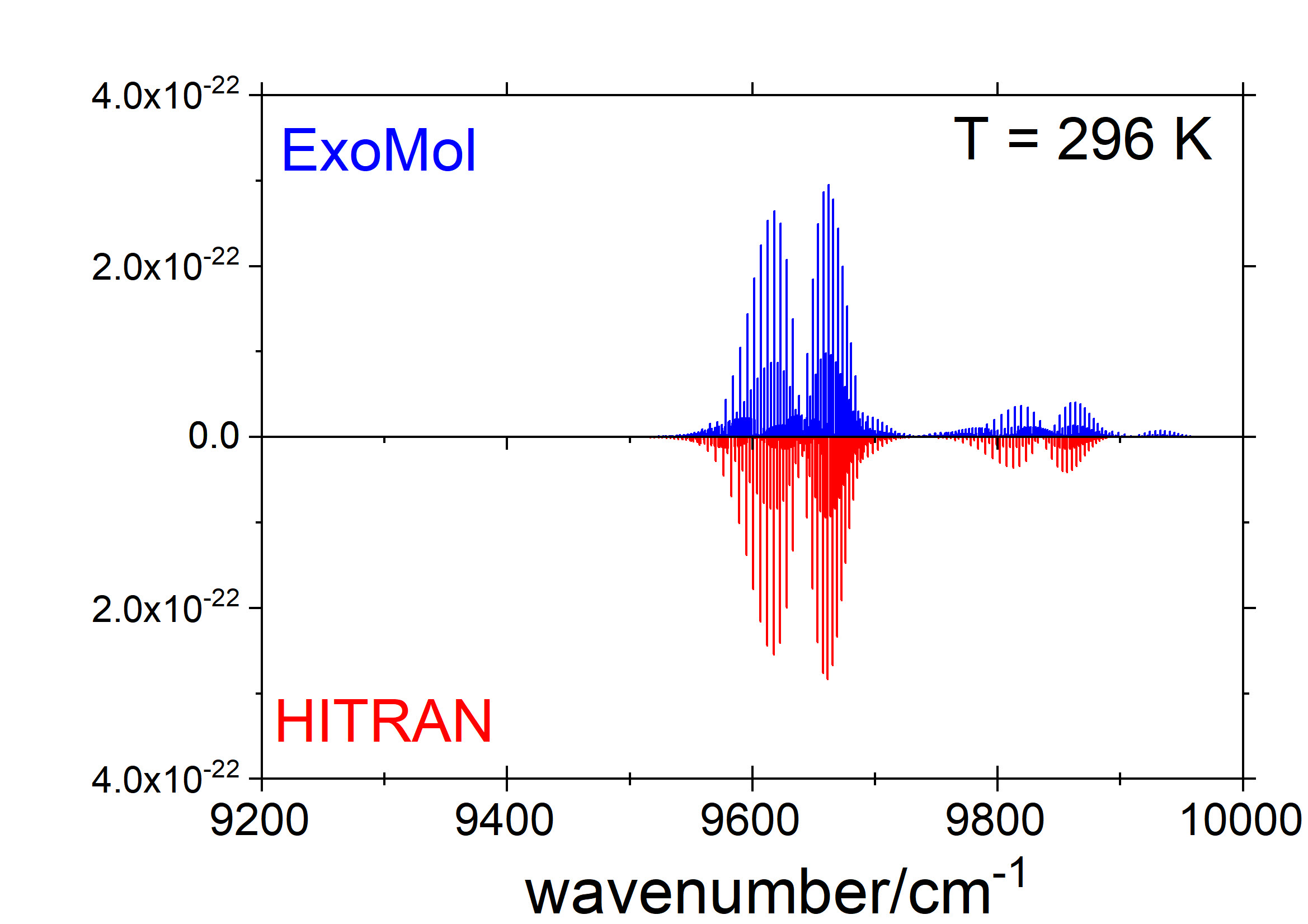}
	\caption{Comparison of the aCeTY stick spectra against HITRAN (with scaling of the dipole moment applied) against HITRAN for different vibrational bands of acetylene at 296~K (in the range 5000-10~000~\cm).  }
	\label{fig:HITRAN:all2}
\end{figure*}

Figure~\ref{fig:HITRAN} shows an overview of an absorption spectrum of \hcch\
computed using aCeTY (this work) to that produced using
HITRAN-2016~\citep{jt691} at $T=$~296~K for the wavenumber range from 0 to
10~000~\cm. Apart from some missing weak bands in HITRAN, it shows a generally
good agreement. Figures~\ref{fig:HITRAN:all} and \ref{fig:HITRAN:all2} give more detailed comparisons of
the main bands of \hcch\ in the range 0--10~000~\cm\ with the data from HITRAN-2016~\citep{jt691}, again in
the form of stick spectra (absorption coefficients) at room-temperature. The
overall agreement of the line positions and intensities is good, except several
weaker bands of \nhcch, overestimated by aCeTY, as well as bands not present in HITRAN.
Table~\ref{t:obs-calc} gives a summary of the accuracy of the line positions for
$J\le 5$ states presented in \Marvel\ and used in the refinement and
band-centre corrections. The full table is given as supplementary information to this work.

\begin{figure}
	\includegraphics[width=\linewidth]{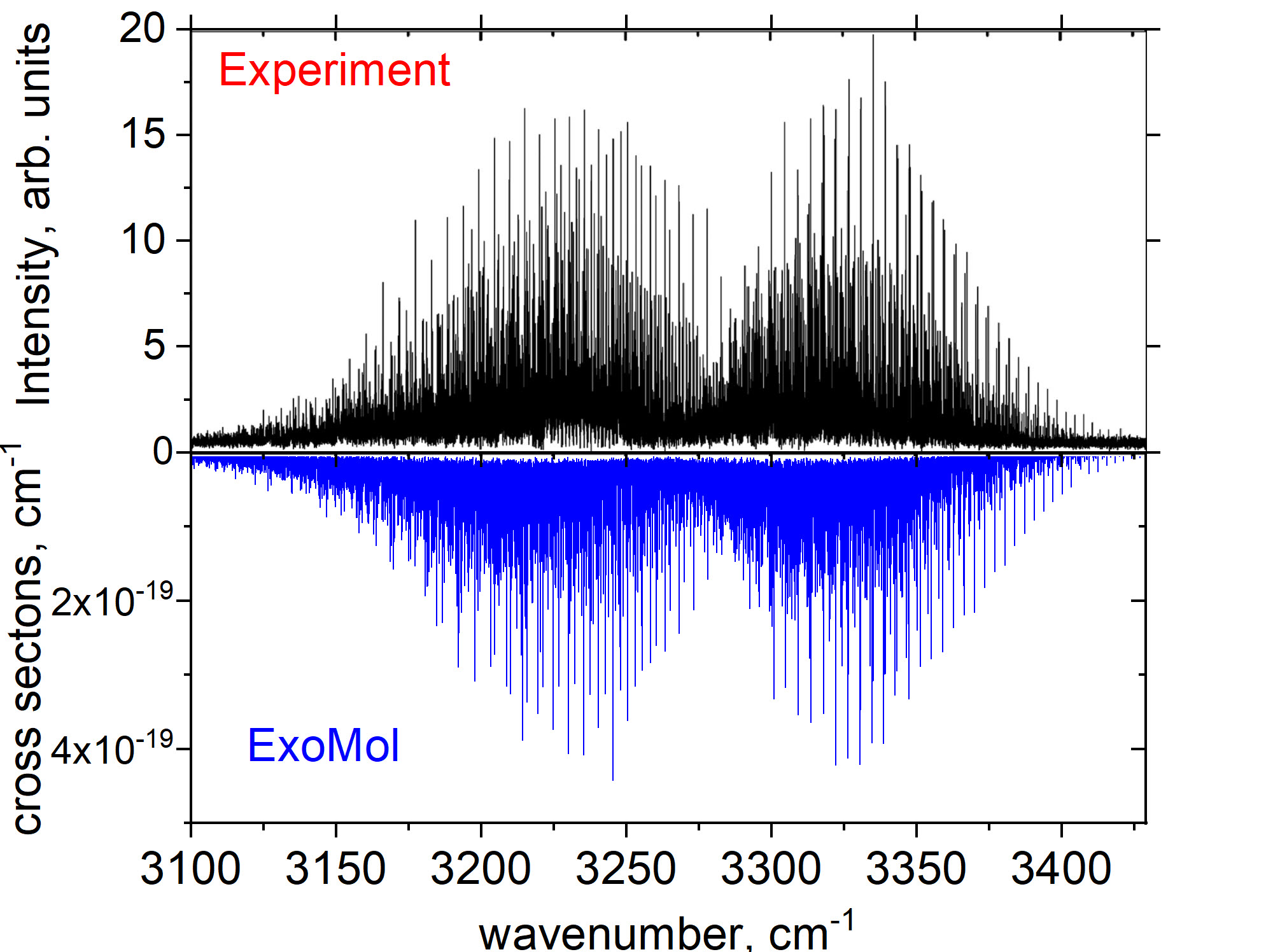}
	\caption{Comparison of the acetylene spectra in the 3$\mu$m region computed using aCeTY at $T=1355$~K with the experimental data by \protect\citet{09AmRoHe.C2H2}. The aCeTY cross-sections were generated using ExoCross and a Voigt line profile assuming $P=1$ atm.}
	\label{fig:3um_zoom}
\end{figure}

Figure~\ref{fig:3um_zoom} shows the hot spectrum (cross-sections) of the
3~$\mu$m band of \nhcch\ at $T=1355$~K compared to the experimental data by
\citet{09AmRoHe.C2H2} demonstrating the generally good agreement also at high
temperatures.

\begin{figure}
	\includegraphics[width=\linewidth]{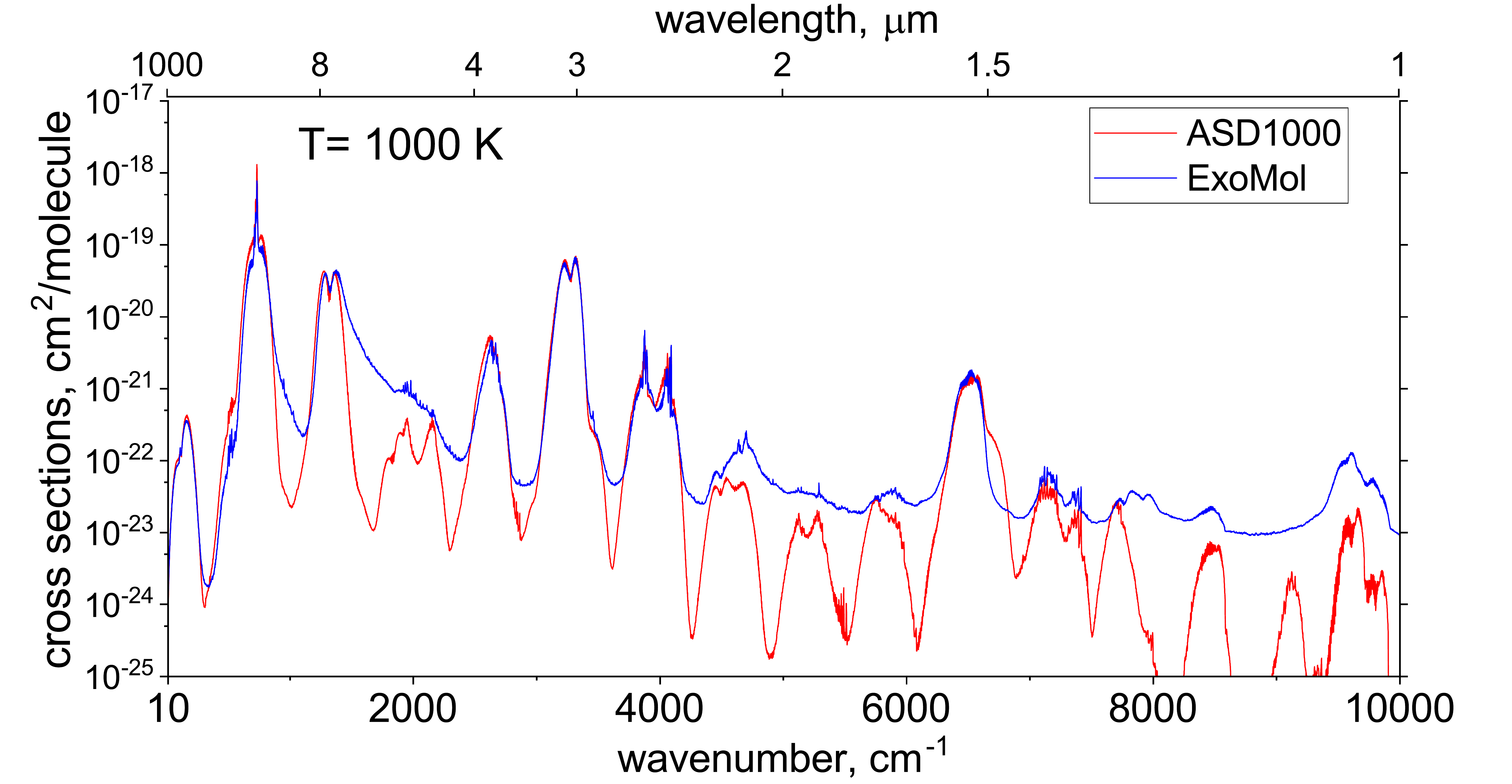}
	\caption{Comparison of the aCeTY line list with ASD-1000~\citep{17LyPe.C2H2}; spectra computed up to 10~000~\cm\ at 1000~K.}
	\label{fig:ASD}
\end{figure}

The
ASD-1000 database of \citet{17LyPe.C2H2} is a calculated acetylene line list  which
covers transitions up 10~000~\cm\ and $J$=100, based on the
use of an effective Hamiltonian fit to experimental data and extrapolated to
higher energies. The energies and intensities at room-temperature agree
reasonably well with those in the HITRAN-2016~\citep{jt691} database (see
\citet{17LyPe.C2H2} and \citet{17LyCaxx.C2H2} for detailed comparisons), and
ASD-1000 has been used to update the 2016 HITRAN release in the low energy
region~\citep{jt691,17JaLyPe.C2H2}. Figure~\ref{fig:ASD} gives a comparison of
cross-sections computed using the aCeTY and ASD-1000 line  lists at $T=1000$~K.
The results are significantly different and it would appear that ASD-1000 fails
to adequately account for the many hot bands which become important at higher temperatures.

\begin{figure}
	\includegraphics[width=\linewidth]{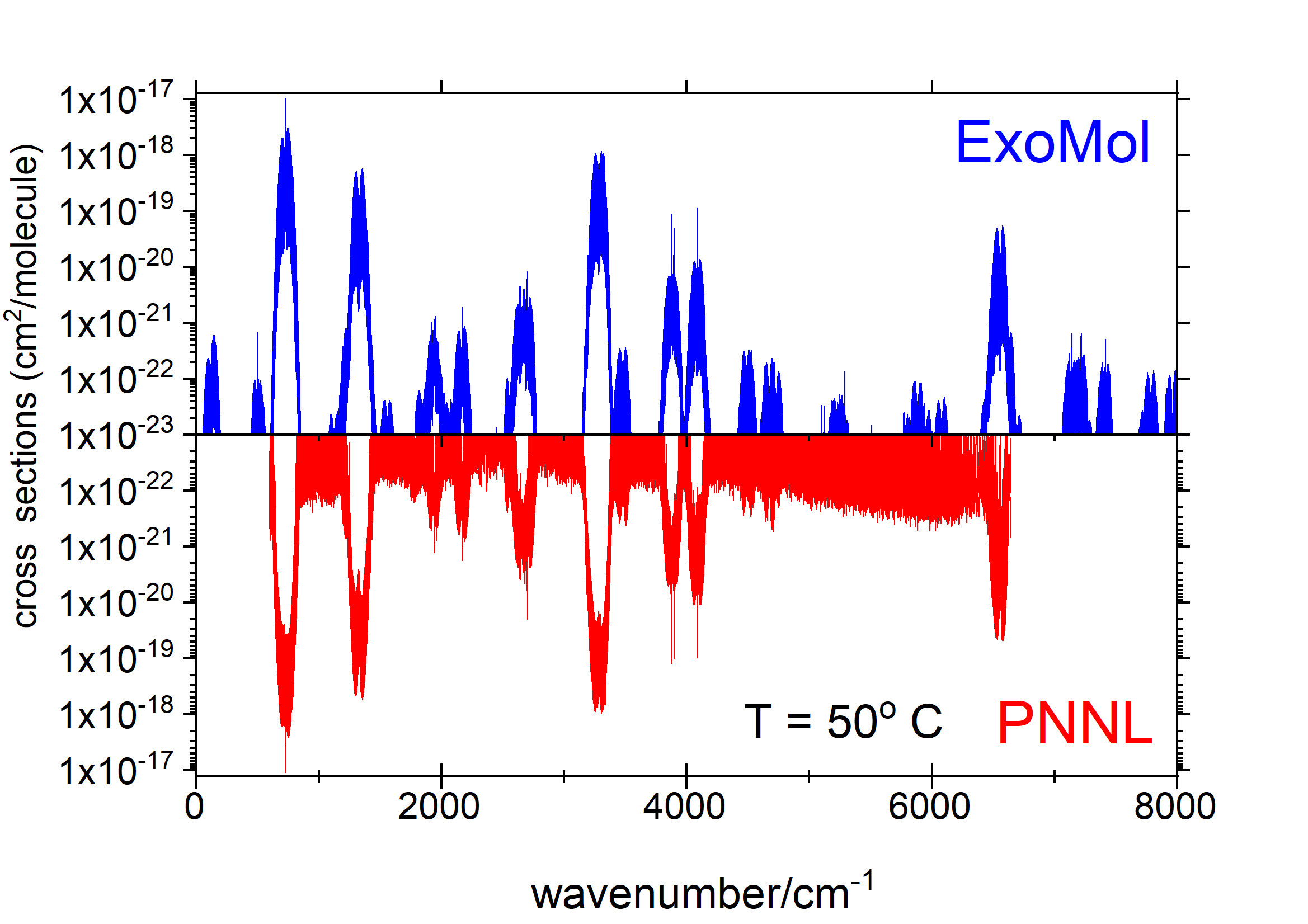}
	\caption{Comparison of the aCeTY cross-sections against PNNL \citep{PNNL} at $T=$~50$^\circ$.
		The aCeTY cross-sections were generated using the Gaussian line profile with half-width-at-half-maximum of $0.01$ \cm.  The PNNL data below $1\times10^{-22}$~cm$^2$~/~molecule are largely due to noise. }
	\label{fig:PNNL}
\end{figure}

A comparison with PNNL is given in Figure~\ref{fig:PNNL}, for
$T=50^{\circ}$.

		\begin{table*}
			\caption{An extract of vibrational transition dipole moment scaling factors, $f_{\mu} = \sqrt{\bar{S}}$, used to produce the line list for fundamental and overtone bands. $\sqrt{\bar{S}}$ is the band intensity scaling factor. The full table is given as part of the supplementary information to this work. }
			\label{t:mu-scale-1}
			\footnotesize
			\begin{tabular}{lrrrrrrrrrr}
				\hline\hline
				$\Gamma$   &$ \upsilon_1$  & $\upsilon_2$ & $\upsilon_3$ & $\upsilon_4$ & $l_4$ & $\upsilon_5$ & $l_5$ & $L$ &   $E_i/hc$ & $f_{\mu}$  \\
				\hline
 $ \Pi_u        $&      0   &      0&   0 &   0 &   0 &   1 &   1 &   1 &   730.33   &      1.0181   \\
 $ \Sigma_u^+   $&      0   &      0&   0 &   1 &   1 &   1 &  -1 &   0 &  1328.07   &      0.8561   \\
 $ \Delta_u     $&      0   &      0&   0 &   1 &   1 &   1 &   1 &   2 &  1347.51   &      0.9813   \\
 $ \Pi_u        $&      0   &      0&   0 &   2 &   2 &   1 &  -1 &   1 &  1941.18   &      0.3653   \\
 $ \Pi_u        $&      0   &      0&   0 &   2 &   0 &   1 &   1 &   1 &  1960.87   &      1.3899   \\
 $ \Pi_u        $&      0   &      0&   0 &   0 &   0 &   3 &   1 &   1 &  2170.34   &      1.0429   \\
 $ \Sigma_u^+   $&      0   &      0&   0 &   3 &   1 &   1 &  -1 &   0 &  2560.59   &      1.7062   \\
 $ \Pi_u        $&      0   &      1&   0 &   0 &   0 &   1 &   1 &   1 &  2703.10   &      1.0271   \\
 $ \Sigma_u^+   $&      0   &      1&   0 &   1 &   1 &   1 &  -1 &   0 &  3281.91   &      1.0268   \\
 $ \Sigma_u^+   $&      0   &      0&   1 &   0 &   0 &   0 &   0 &   0 &  3294.85   &      0.8923   \\
 $ \Pi_u        $&      0   &      1&   0 &   2 &   0 &   1 &   1 &   1 &  3882.42   &      0.8122   \\
 $ \Pi_u        $&      0   &      0&   1 &   1 &   1 &   0 &   0 &   1 &  3898.34   &      0.5774   \\
 $ \Pi_u        $&      1   &      0&   0 &   0 &   0 &   1 &   1 &   1 &  4092.34   &      0.9985   \\
 $ \Pi_u        $&      0   &      1&   0 &   0 &   0 &   3 &   1 &   1 &  4140.08   &      0.9636   \\
 $ \Sigma_u^+   $&      0   &      1&   0 &   3 &   1 &   1 &  -1 &   0 &  4488.85   &      0.6464   \\
 $ \Sigma_u^+   $&      0   &      0&   1 &   2 &   0 &   0 &   0 &   0 &  4508.02   &      0.7419   \\
 $ \Sigma_u^+   $&      1   &      0&   0 &   1 &   1 &   1 &  -1 &   0 &  4673.71   &      1.5789   \\
 $ \Sigma_u^+   $&      0   &      0&   1 &   0 &   0 &   2 &   0 &   0 &  4727.07   &      0.8830   \\
				\hline
				\hline
			\end{tabular}
		\end{table*}

		\begin{table*}
			\caption{An extract of the transition dipole moment scaling factors, $f_{\mu} = \sqrt{\bar{S}}$, used to produce the line list for hot bands starting from the
				$(000100)$ $\Pi_g$ state. $\sqrt{\bar{S}}$ is the band intensity scaling factor. The full table is given as part of the supplementary information to this work. }
			\label{t:mu-scale-2}
			\footnotesize
			\begin{tabular}{lrrrrrrrrrr}
				\hline\hline
				$\Gamma$   &$ \upsilon_1$  & $\upsilon_2$ & $\upsilon_3$ & $\upsilon_4$ & $l_4$ & $\upsilon_5$ & $l_5$ & $L$ &  $E_i/hc$ & $f_{\mu}$  \\
				\hline
 $ \Pi_u        $&      0   &      0&   0 &   0 &   0 &   1 &   1 &   1 &   730.33   &      0.8730   \\
 $ \Sigma_u^+   $&      0   &      0&   0 &   1 &   1 &   1 &  -1 &   0 &  1328.07   &      1.0490   \\
 $ \Delta_u     $&      0   &      0&   0 &   1 &   1 &   1 &   1 &   2 &  1347.51   &      1.0312   \\
 $ \Pi_u        $&      0   &      0&   0 &   2 &   2 &   1 &  -1 &   1 &  1941.18   &      0.8233   \\
 $ \Pi_u        $&      0   &      0&   0 &   2 &   0 &   1 &   1 &   1 &  1960.87   &      1.5482   \\
 $ \Sigma_u^+   $&      0   &      0&   0 &   3 &   1 &   1 &  -1 &   0 &  2560.59   &      0.5440   \\
 $ \Delta_u     $&      0   &      0&   0 &   3 &   3 &   1 &  -1 &   2 &  2561.67   &      0.4387   \\
 $ \Sigma_u^-   $&      0   &      0&   0 &   3 &   1 &   1 &  -1 &   0 &  2583.84   &      1.3237   \\
 $ \Pi_u        $&      0   &      1&   0 &   0 &   0 &   1 &   1 &   1 &  2703.10   &      0.7272   \\
 $ \Sigma_u^+   $&      0   &      0&   0 &   1 &   1 &   3 &  -1 &   0 &  2757.80   &      1.2130   \\
 $ \Delta_u     $&      0   &      0&   0 &   1 &   1 &   3 &   1 &   2 &  2773.41   &      1.0312   \\
 $ \Sigma_u^-   $&      0   &      0&   0 &   1 &   1 &   3 &  -1 &   0 &  2783.63   &      0.9999   \\
 $ \Delta_u     $&      0   &      0&   0 &   1 &  -1 &   3 &   3 &   2 &  2796.30   &      1.1276   \\
 $ \Sigma_u^+   $&      0   &      1&   0 &   1 &   1 &   1 &  -1 &   0 &  3281.91   &      0.8752   \\
 $ \Sigma_u^+   $&      0   &      0&   1 &   0 &   0 &   0 &   0 &   0 &  3294.85   &      0.8065   \\
 $ \Pi_u        $&      0   &      1&   0 &   2 &   0 &   1 &   1 &   1 &  3882.42   &      1.1363   \\
 $ \Pi_u        $&      0   &      0&   1 &   1 &   1 &   0 &   0 &   1 &  3898.34   &      0.8653   \\
 $ \Sigma_u^+   $&      1   &      0&   0 &   1 &   1 &   1 &  -1 &   0 &  4673.71   &      0.9740   \\
 $ \Sigma_u^-   $&      1   &      0&   0 &   1 &   1 &   1 &  -1 &   0 &  4688.83   &      1.0084   \\
 $ \Delta_u     $&      1   &      0&   0 &   1 &   1 &   1 &   1 &   2 &  4692.06   &      0.9710   \\
				\hline
				\hline
			\end{tabular}
		\end{table*}

		\begin{table*}
			\caption{Vibrational transition dipole moment scaling factors, $f_{\mu} = \sqrt{\bar{S}}$, used to produce the line list: hot bands starting from the
				$(0000011)$ $\Pi_u$ state. $\sqrt{\bar{S}}$ is the band intensity scaling factor. }
			\label{t:mu-scale-3}
			\footnotesize
			\begin{tabular}{lrrrrrrrrrrr}
				\hline\hline
				$\Gamma$   &$ \upsilon_1$  & $\upsilon_2$ & $\upsilon_3$ & $\upsilon_4$ & $l_4$ & $\upsilon_5$ & $l_5$ & $L$  &  $E_i/hc$ & $f_{\mu}$  \\
				\hline
 $ \Sigma_g^+   $&      0   &      0&   0 &   2 &   0 &   0 &   0 &   0 &  1230.38   &      1.3200   \\
 $ \Delta_g     $&      0   &      0&   0 &   2 &   2 &   0 &   0 &   2 &  1233.49   &      0.6750   \\
 $ \Sigma_g^+   $&      0   &      0&   0 &   0 &   0 &   2 &   0 &   0 &  1449.11   &      1.0060   \\
 $ \Delta_g     $&      0   &      0&   0 &   0 &   0 &   2 &   2 &   2 &  1463.00   &      0.9773   \\
 $ \Sigma_g^+   $&      0   &      1&   0 &   0 &   0 &   0 &   0 &   0 &  1974.35   &      1.0368   \\
 $ \Pi_g        $&      0   &      0&   0 &   1 &   1 &   2 &   0 &   1 &  2049.06   &      0.8462   \\
 $ \Pi_g        $&      0   &      0&   0 &   1 &  -1 &   2 &   2 &   1 &  2066.97   &      0.7603   \\
 $ \Pi_g        $&      0   &      1&   0 &   1 &   1 &   0 &   0 &   1 &  2574.76   &      0.8972   \\
 $ \Sigma_g^+   $&      0   &      0&   0 &   2 &   2 &   2 &  -2 &   0 &  2648.01   &      0.3475   \\
 $ \Sigma_g^-   $&      0   &      0&   0 &   2 &   2 &   2 &  -2 &   0 &  2661.16   &      0.4196   \\
 $ \Delta_g     $&      0   &      0&   0 &   2 &   2 &   2 &   0 &   2 &  2666.06   &      0.5947   \\
 $ \Sigma_g^+   $&      0   &      0&   0 &   0 &   0 &   4 &   0 &   0 &  2880.22   &      1.0651   \\
 $ \Delta_g     $&      0   &      0&   0 &   0 &   0 &   4 &   2 &   2 &  2894.04   &      1.0543   \\
 $ \Sigma_g^+   $&      1   &      0&   0 &   0 &   0 &   0 &   0 &   0 &  3372.84   &      0.9953   \\
 $ \Pi_g        $&      1   &      0&   0 &   1 &   1 &   0 &   0 &   1 &  3970.05   &      1.5178   \\
 $ \Pi_g        $&      0   &      1&   0 &   1 &   1 &   2 &   0 &   1 &  4002.46   &      1.2100   \\
 $ \Pi_g        $&      0   &      0&   1 &   0 &   0 &   1 &   1 &   1 &  4016.73   &      0.8622   \\
 $ \Sigma_g^+   $&      1   &      0&   0 &   0 &   0 &   2 &   0 &   0 &  4800.13   &      0.9912   \\
 $ \Delta_g     $&      1   &      0&   0 &   0 &   0 &   2 &   2 &   2 &  4814.19   &      0.9952   \\
				\hline
				\hline
			\end{tabular}
		\end{table*}

		\section{Exoplanet Atmospheres}\label{sec:exo}
		
\begin{figure}
		\includegraphics[width=\linewidth]{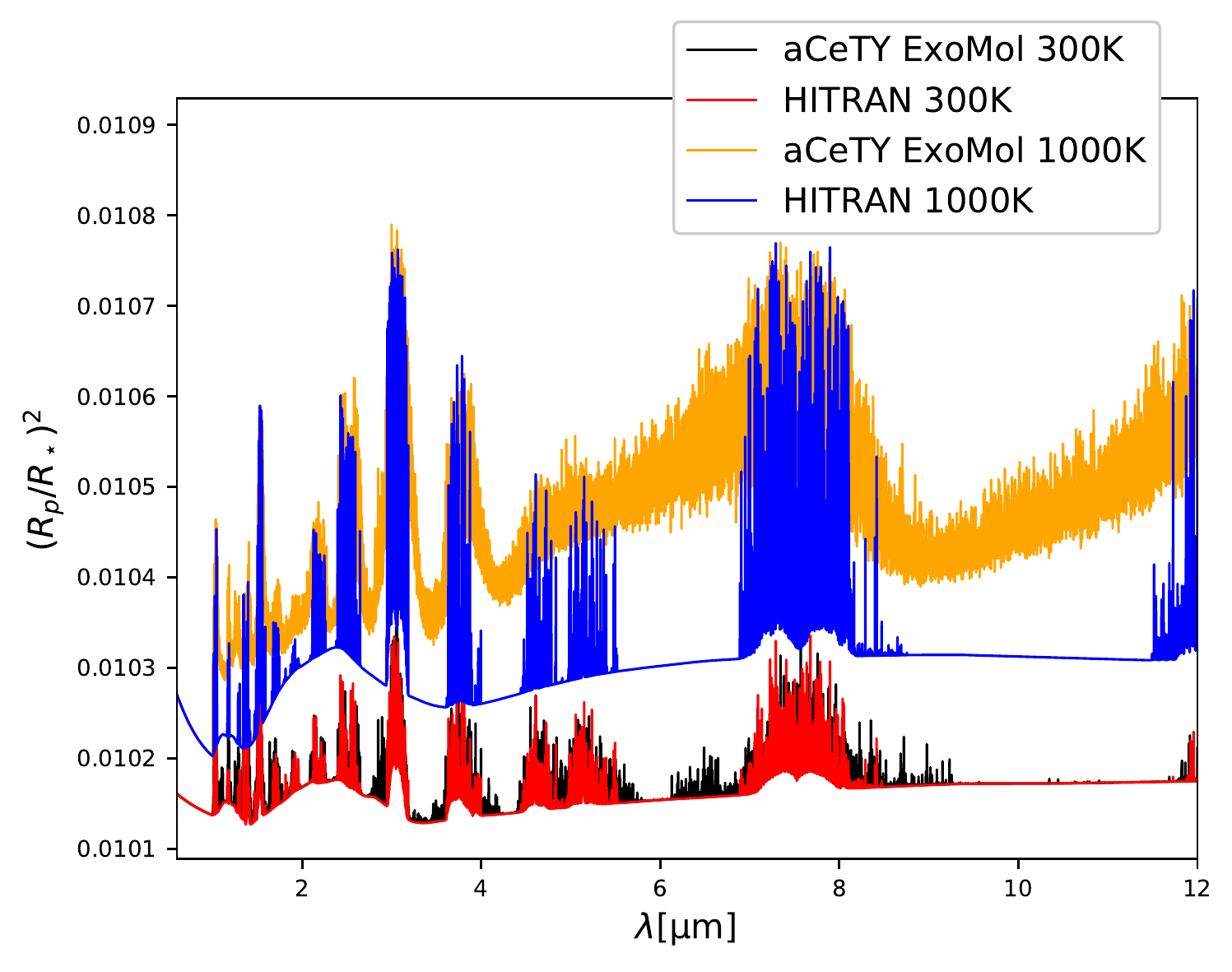}
		\caption{The transmission spectra of a hypothetical planetary atmosphere of a Jupiter-size planet around a solar-like star, with an atmosphere of pure \hcch, at 1000~K, computed using TauREx~\citep{15WaTiRo.exo}. A comparison is made using aCeTY  (this work) against HITRAN line list data as input into the cross-sections used in the transmission spectrum computation.}
		\label{fig:taurex}
	\end{figure}

\begin{figure}
			\includegraphics[width=\linewidth]{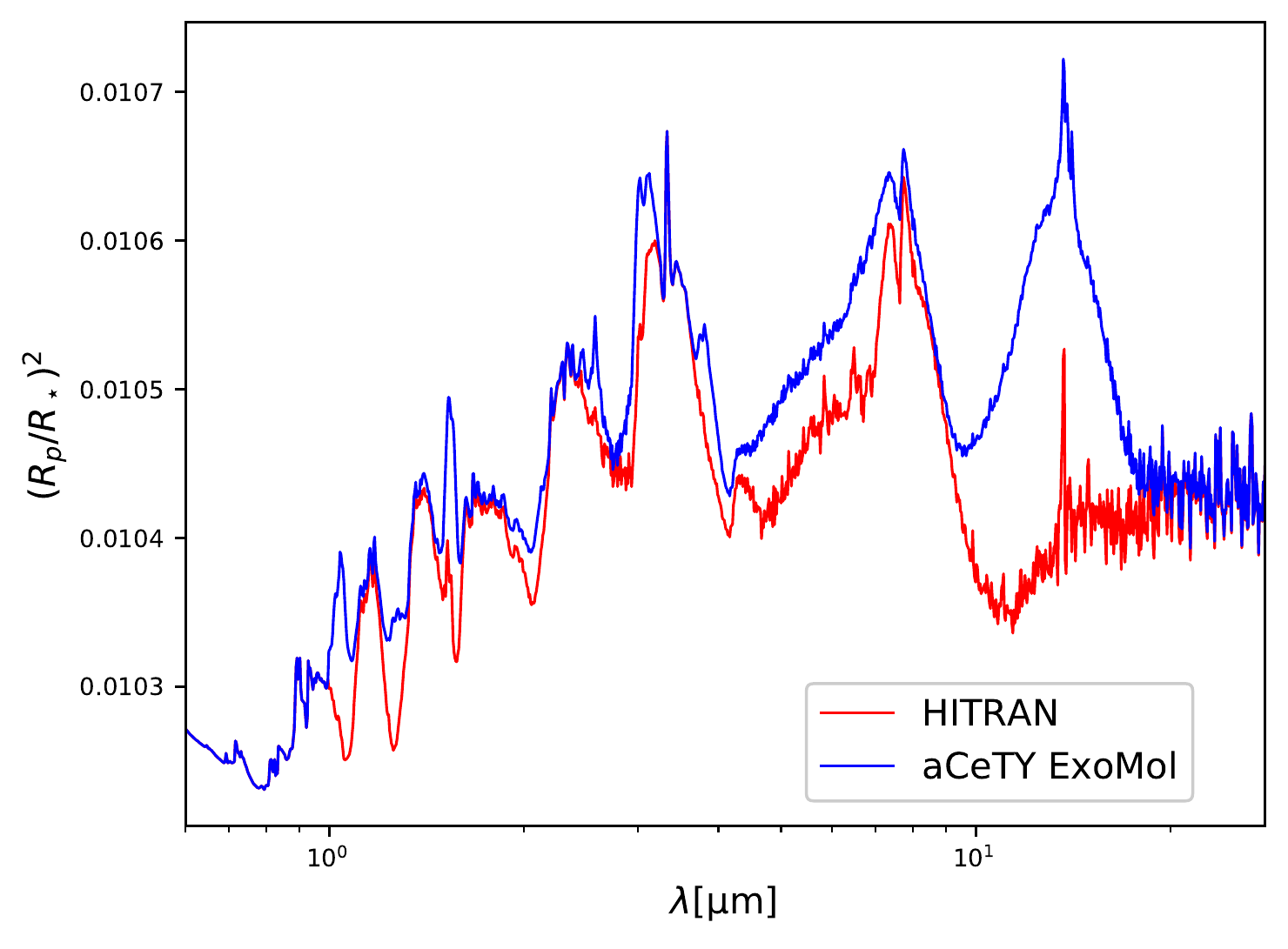}
			\caption{The transmission spectra of a hypothetical planetary atmosphere of a Jupiter-size planet around a solar-like star, with an atmosphere of H$_2$O, CO$_2$, CH$_4$, CO, HCN, H$_2$S and C$_2$H$_2$ at approximately equilibrium abundances at 1000~K, computed using TauREx~\citep{15WaTiRo.exo}. A comparison is made using aCeTY (this work) against HITRAN line list data as input into the transmission spectrum computation for C$_2$H$_2$. All other molecules and parameters remain the same between the two spectra. }
			\label{fig:taurex2}
		\end{figure}

		Figure~\ref{fig:taurex} gives the transmission spectra of a hypothetical planetary atmosphere of a Jupiter-size planet around a solar-like star, with an atmosphere of pure \hcch, at 1000~K, computed using TauREx~\citep{15WaTiRo.exo}. A comparison of such an atmosphere using the aCeTY line list is made against one using HITRAN cross-section data (both are computed at a resolving power of R=${\lambda}/{\Delta\lambda}=$10~000). It can be seen that a large amount of opacity would be lost if one used HITRAN data at high-temperatures. This is further demonstrated by an atmosphere of a planet with the same mass and temperature, which contains
		 H$_2$O~\citep{jt734}, CO$_2$~\citep{HITRAN}, CH$_4$~\citep{jt698}, CO~\citep{15LiGoRo.CO}, HCN~\citep{jt570}, H$_2$S~\citep{jt640}
		and C$_2$H$_2$ at approximately equilibrium abundances. Figure~\ref{fig:taurex2} shows the differences between using aCeTY and HITRAN data as input into the transmission spectrum computation for C$_2$H$_2$. Low resolution (R=${\lambda}/{\Delta\lambda}=$300) k-tables are used here. All other molecules and parameters remain the same between the two spectra. The cross-sections and k-tables (the latter are produced using a method of opacity sampling which enables low resolution computations while still taking strong opacity fluctuations at high resolution into account; see, for example, \cite{17Min.methods}) used in these model will shortly be made publicly available  \citep{19Chubbetal}.

\section{Summary and Future Work}\label{summary}

		In this work we present a new ro-vibrational line list for the
ground electronic state of the main isotopologue of acetylene, \hcch; the aCeTY
line list. This line list was computed as part of the ExoMol project
\citep{jt528,jt631}, for characterising exoplanet and cool stellar atmospheres.
It is considered complete up to 2200~K, with transitions computed up to
10~000~\cm\  (down to 1~$\mu$m), with lower and upper energy levels up to
12~000~\cm\ and 22~000~\cm\ considered, respectively.
		The calculations were performed up to a maximum value for the
vibrational angular momentum, $K_{\rm max}=L_{\rm max}$~=~16, and maximum
rotational angular momentum, $J$~=~99. The aCeTY line list is based on \ai\
electronic structure calculations for the potential energy and dipole moment
surfaces, but with improvements on the accuracy of both the line positions and
the dipole moments made using the wealth of experimental data available from the
literature.

Comparisons against other available line list data
demonstrate that the aCeTY line list is the most complete and accurate available
line list for acetylene to date. It is therefore recommended for use in
characterising exoplanet and cool stellar atmospheres. Computing cross-section
and k-table opacity data for \hcch, at a range of temperatures and pressures
suitable for use in exoplanet atmospheres, for use in retrieval codes such as
Tau-REx~\citep{15WaTiRo.exo}, ARCiS~\citep{19MiOrCh.arcis},
NEMESIS~\citep{NEMESIS} and petitRADTRANS~\citep{19MoWaBo.petitRADTRANS} will be
published in soon~\citep{19Chubbetal}.

For high-resolution applications,
however, some caution is advised. More work needs to be done, which is
ongoing, to further improve and ensure the accuracy of line position
to achieve  precision required for studies of exoplanets using high resolution  Doppler spectroscopy.
The line list can be improved by augmenting the \Marvel\ analysis of \cite{jt705}
with new laboratory data, such as \cite{18TwHaSe.C2H2,19DiAiDe.HCCH,19NuAlCh.C2H2},
and then inserting the resulting
energy levels in aCeTY; this would ensure that energies and associated transition wavenumbers are at the current limit of accuracy. The
high-accuracy experiments of \cite{18tAHuSu.C2H2} and \cite{13LiLiWa.C2H2}
(which was not included in \cite{jt705}) demonstrates that updates should be made to
the $\nu_1+3\nu_3$ band included in the \Marvel\ analysis of \hcch.

The new intensity scaling technique presented in this work will be useful for future high-precision spectroscopic applications, especially if combined with the MARVELisation procedure. It has the potential to target the accuracy of experiment when predicting line intensities within a given vibrational band at different temperatures.

The line lists aCeTY can be downloaded from the CDS, via
\url{ftp://cdsarc.u-strasbg.fr/pub/cats/J/MNRAS/}, or
\url{http://cdsarc.u-strasbg.fr/viz-bin/qcat?J/MNRAS/}, or from
\url{www.exomol.com}.

\section*{Acknowledgements}

This work was supported by the UK Science and Technology Research Council (STFC) No.
ST/R000476/1 and through a studentship to KLC. This work made extensive use of UCL's Legion high
performance computing facility along with the STFC DiRAC HPC facility supported by BIS National
E-infrastructure capital grant ST/J005673/1 and STFC grants ST/H008586/1 and ST/K00333X/1.
KLC acknowledges funding from the European Union's Horizon 2020 Research and Innovation Programme, under Grant Agreement 776403.
		
		
		


		\label{lastpage}
	\end{document}